\documentclass[
preprint,
aps,superscriptaddress,floats,showpacs,pdftex]{revtex4}
\usepackage{graphicx}
\usepackage{amsmath}
\usepackage{amssymb}
\usepackage{epsfig}

\begin{document}

\title{Relativistic Laser-Matter Interaction and Relativistic Laboratory
Astrophysics}
\author{S. V. Bulanov$^{1,2}$, T. Zh. Esirkepov$^1$, D. Habs$^{3,4}$, F.
Pegoraro$^5$, T. Tajima$^{1,3,4}$\\
$^1${\small {Advanced Photon Research Center, JAEA, 8-1 Umemidai,
Kizugawa, 619-0215 Kyoto, Japan}}\\
$^2${\small {A. M. Prokhorov Institute of General Physics, RAS, Vavilov
street, 38, 119991 Moscow, Russia}}\\
$^3${\small {Sektion Physik, Ludwig-Maximilians-Universitaet Muenchen,
D-85748 Carching, Germany}}\\
$^4${\small {Max-Planck-Institut fuer Quantenoptik, D-857748 Garching,
Germany}\\ 
$^5${Physics Dept. and CNISM, University of Pisa, Largo Pontecorvo,
3, 56127 Pisa, Italy} }
}
\begin{abstract}
The paper is devoted to the prospects of using the laser radiation
interaction with plasmas in the laboratory relativistic astrophysics
context. We discuss the dimensionless parameters characterizing the
processes in the laser and astrophysical plasmas and emphisize a similarity
between the laser and astrophisical plasmas in the ultrarelativistic energy
limit. In particular, we address basic mechanisms of the charged particle
acceleration, the collisionless shock wave and magnetic reconnection and
vortex dynamics properties relevant to the problem of ultrarelativistic
particle acceleration.
\end{abstract}

\pacs{52.27.Ny, 52.72.+v}
\maketitle
\newpage
\tableofcontents
\newpage
\section{Introduction}

\label{intro}

High-power laser facilities have made unprecedented progress in recent years
and the nearest future their radiation may reach intensities of 10$^{24}$W/cm%
$^{2}$ and higher \cite{ELIILE}. As a result of laser technology progress
the laser-matter interaction entered regimes of interest for astrophysics.
Typically in the course of laser irradiation of targets shock waves are
generated; the target compression is accompanied by the Rayleigh-Taylor (RT)
and Richtmayer-Meshkov (RM) instability development; collimated plasma jets
are observed; the matter equation of state (EOS) acquires new properties
under extreme pressure, density and temperature conditions; the laser plasma
emits high energy charged particle beams and high- and low-frequency
powerful electromagnetic radiation. Gathering of these facts principal for
both space and laboratory physics has initiated works in the so-called
laboratory astrophysics \cite{LabAstr} with the aim to model the processes
of key importance for the space objects under laboratory conditions.
Concerning the laser facilities, the present day laser systems can be
subdivided into two categories. The first category includes lasers with a
relatively long pulse of pico- and nanosecond duration and generally low
repetition rate. These high energy and power laser facilities have been
mainly developed for purposes of inertial confinement fusion with the laser
pulse and target parameters corresponding to the collisional hydrodynamics
phenomena \cite{Lindl}. In context of laboratory astrophysics they are used
for experiments on shock waves, including the radiative shocks and RT\&RM
instability, the jet formation, and the EOS studies. The second category
includes table top size lasers, whose pulse duration is of the order of a
few tens of a femto-second with high repetition rate \cite{S-Mou}. Due to
ultra short pulse duration and high contrast, these relatively moderate
energy lasers can produce extremely high power and relativistically high
intensity electromagnetic pulses. However, the role of both kinds of laser
systems is complementary for the development of experimental facilities for
the purposes of relativistic laboratory astrophysics.

Generic questions for astrophysics such as whether we are living in the
Universe or in the Multiverse \cite{Weinberg}, related discussions of the
inflation era in the Multiverse evolution \cite{Linde} and probing our
world's dimensions are related to quantum gravitation physics and deal with
the observational cosmology, in particular with an analysis of the cosmic
black body radiation, the nuclear synthesis and Type I supernovae radiation
(see \cite{Khlopov}), are yet out of the energy range accessible with
present day lasers. The quantum gravitation energy scale is given by the
Planck energy, $\sqrt{\hbar c^{3}/G}\approx 10^{19}$GeV, which corresponds
to the mass $\sqrt{\hbar c/G}\approx 10^{-5}$g and the length $\sqrt{\hbar
G/c^{3}}\approx 10^{-33}\ $cm. In quantum field theory the unification
energy scale corresponds to $10^{16}$GeV \cite{Perkins,Wilczek}. These
energy frontiers are yet well above of nowadays laser pulse energies.
Fortunately, new physics such as the Higgs boson detection and exploration
of the physics beyond the Standard Model is anticipated to be met at a
substantially lower energy level in the range of several TeV in the
experiments planned with the Large Hadron Collider (LHC), as summarized in
Ref. \cite{Wilczek}. If relativistic laser plasmas can provide the charged
particle acceleration up to the TeV energy level, laser accelerators will
make a considerable impact to high energy physics, to finding answers on
black hole and brane production under the terrestrial conditions \cite{MBH},
to test causality \cite{TM} and to study the quark-gluon plasmas \cite{JR}.

We may see that the main field of studies of astrophysical phenomena with
high power lasers lies in the electrodynamics of continuous media in the
relativistic regime \cite{MTB-06}. Since matter irradiated by ultrastrong
electromagnetic waves (EMW) is ionized during a time interval comparable
with the wave period and becomes a plasma and under astrophysical conditions
approximately 95\% of barionic part of matter is in the plasma state, the
object of our studies is the relativistic laser and astrophysical plasma.

If we address to the problems of contemporary relativistic astrophysics,
first of all questions on the mechanisms of the cosmic ray acceleration and
on the properties of strong EMW interaction with relativistic plasmas
attract our attention \cite{ACR}. In space plasmas basic mechanisms of
charged particle acceleration are connected with the reconnection of
magnetic field lines, which is accompanied by the strong and regular
electric field generation (it occurs in the planet magnetosheres, in binary
stellar systems, in accretion disks, in the magnetar magnetospheres, etc.)
and with collisionless shock waves, at the fronts of which the charged
particle acceleration occurs (this happens in interplanetary space, during
supernova explosions, in colliding galaxies, etc.) \cite{ACR,AH}.

The laser accelerator development relies upon the fact that under the
terrestrial laboratory conditions presently one of the most powerful sources
of coherent electromagnetic radiation is provided by lasers \cite{MTB-06}.
Wakefield accelerators, \cite{T-D} and \cite{Ch-D}, presently provide the
most advanced schemes for electron acceleration and they may be suggested to
be good candidates for the charged particle acceleration in space \cite%
{ACR,CTT}. One of the efficient mechanisms of ion acceleration in laser
plasmas utilizes the radiation pressure of electromagnetic waves interacting
with plasmas (see Refs. \cite{RPDA} and \cite{BEKT} ). Radiation pressure is
a very effective mechanism of momentum transfer to charged particles. This
mechanism was introduced long ago \cite{leb} and physical conditions of
interest range from stellar structures and radiation generated winds (see
e.g. Refs. \cite{miln}), to the formation of \textquotedblleft photon
bubbles\textquotedblright \ in very hot stars and accretion disks \cite%
{Arons}, to particle acceleration in the laboratory \cite{RPDA,VEK,TER}, see
in addition Refs. \cite{lif,ma}, and in high energy astrophysical
environments \cite{astr1}.

Utilization of the plasma nonlinear properties for the electromagnetic wave
intensification can result in much higher intensity and power. In this case
a fundamental role is played by relativistic mirrors, which are thin
electron sheets induced by the laser radiation moving with a speed close to
the speed of light in vacuum, as proposed in Ref. \cite{BET-03}. We note the
fruitfulness of the relativistic mirror concept for solving a wide range of
problems in modern theoretical physics. Relativistic mirrors are important
elements in the theory of the dynamical Casimir effect \cite{Casimir}, with
regard to the Unruh radiation \cite{Unruh} and other nonlinear vacuum
phenomena \cite{Rozanov,four-wave,Narozhny-Fedotov,MSh-06,SHHK-06}.
Relativistic mirrors made by wake waves may lead to an electromagnetic wave
intensification resulting in an increase of pulse power up to the level when
the electric field of the wave reaches the Schwinger limit \cite{SCHWIN}
when electron-positron pairs are created from the vacuum and the vacuum
refractive index becomes nonlinearly dependent on the electromagnetic field
strength. In quantum field theory particle creation from the vacuum attracts
a great attention, because it provides a typical example of non perturbative
processes \cite{QED}. Nonlinear QED vacuum properties can in future be
probed with such strong and powerful electromagnetic pulses.

If we trace a relationship between astrophysics and laser physics, we can
see a number of publications devoted to the laboratory modeling of
astrophysical processes \cite{LabAstr}. As known there has been an interest
in modeling space physics with laboratory experiments for many years. The
first modeling of processes fundamental for space physics in terrestrial
laboratories has been done by Kristian Birkeland, who more than 100 years
ago conducted first experiments on studying the auroral regions in the earth
magnetosphere \cite{BIRK}. Lateron progress has been achieved in the
laboratory modeling of various processes \cite{SPEXP,FORT}, including the
magnetic field reconnection \cite{REC}, collisionless shock waves \cite{SHW}%
, which provide mechanisms for charged particle acceleration under various
astrophysical conditions (see Ref. \cite{ACR}).

In the present paper we address plasma processes relevant to space physics,
which occur in the relativistic and collisionless regimes.

\section{Dimensionless Parameters that Characterize the Interaction Regimes
of High Intensity Electromagnetic Waves with Matter}

\subsection{Principle of Qualitative Scaling}

Laboratory experiments for studying astrophysical phenomena are of two types 
\cite{FH}. The first type of experiments can be referred to as \textit{%
configuration modeling}, which is aiming at simulating the actual
configuration of a system, e.g. the whole Earth's magnetosphere (for example
see Ref. \cite{ZABM}, where the results of the laser-plasma experiments on
the simulation of the global impact of the coronal mass ejections onto the
Earth's magnetosphere are presented). The second type of experiments
corresponds to \textit{process simulation}, i. e. they are aiming at
studying the properties of physical processes relevant to astrophysical
phenomena \cite{LabAstr}. There are a number of nonlinear plasma physical
processes that require their clarification.

Physical systems obey scaling laws, which can also be presented as
similarity rules. In the theory of similarity and modeling the key role is
played by dimensionless parameters that characterize the phenomena under
consideration \cite{Sedov}. The principle requirement of the laboratory
modeling is the equality of the key dimensionless parameters in the modeled
processes. In cases of modeling astrophysical phenomena where this equality
can hardly be respected, instead the\textit{\ principle of limited similarity%
} (PLS) or \textit{principle of qualitative scaling} has been formulated in
Refs. \cite{PodSag} and \cite{FH}. According to the PLS those dimensionless
parameters, which are relevant in a certain context and which are much
larger or smaller than unity under astrophysical conditions must retain this
property (i.e. be much larger or smaller than unity) in the laboratory
experiments modeling the astrophysical process. Below we present the key
dimensionless parameters that characterize the high intensity
electromagnetic wave (EMW) interaction with matter (see also Refs. \cite%
{NEW,RELPEG}).

\subsection{\protect\smallskip Parameters of Strong EMW Propagating in
Plasmas}

The intensity of an electromagnetic wave pulse is defined by its electric
field amplitude through the expression: $I=cE_{0}^{2}/4\pi $, which is
related to the Poynting vector 
\begin{equation}
P=\frac{c}{4\pi }\left[ {E\times B}\right] .
\end{equation}
The power of the EMW is equal to the integral over its transverse
cross-section $S$, 
\begin{equation}
P=\frac{{c}}{4\pi }\oint\limits_{S}{\left( {\left[ {E\times B}\right] \cdot n%
}\right) }dS=I\,S.
\end{equation}
The time integral of the power gives the pulse energy, $\mathcal{E}=\mathcal{%
P}\tau _{p}$, where $\tau _{p}$, is the pulse duration. Other important
parameters are the pulse frequency, $\omega _{0}$, which is related to its
wavelength, $\lambda _{0}=2\pi c/\omega _{0}$, and the pulse polarization.

The first of the dimensionless parameters which characterizes the EMW packet
is the ratio of the pulse length, $l_{p}=c\tau _{p}$, to the radiation
wavelength, $\lambda _{0}$. We shall denote this ratio as $%
N_{p}=l_{p}/\lambda _{0}$. It is equal to the number of wavelengths per
pulse, and is Lorentz invariant.

If the EMW intensity is relatively low, irradiated matter is not ionized. We
notice that the typical energy of a photon in the laser parameter range with
wavelengths in the micron range is of the order of one electron-volt and is
substantially smaller than the binding energy of an electron inside an atom, 
$\hbar \omega _{0}\ll W_{b}$, i.e. it is smaller than the atomic ionization
potential. In this case the characteristic dimensionless parameter of the
interaction is the ratio between the amplitude of the electric field in the
laser pulse, $E=\sqrt{4\pi I/c}$, and the atomic electric field, $E_{a}$.
The latter is equal to the electric field of the proton at a distance of a
Bohr radius, $a_{B}=\hbar ^{2}/m_{e}e^{2}\simeq 5.3\times 10^{-9}cm$, i.e. $%
E_{a}=e/a_{B}^{2}=m_{e}^{2}e^{5}/\hbar ^{4}$. The electron binding energy is 
$W_{b}=\hbar ^{2}/2a_{B}^{2}m_{e}$ and it corresponds to the frequency $%
\omega _{a}=W_{b}/\hbar $. The above condition, $\hbar \omega _{0}\ll W_{b}$%
, is equivalent to the inequality $\hbar \omega _{0}/W_{b}=\omega
_{0}/\omega _{a}\ll 1$. The dimensionless parameter 
\begin{equation}
\frac{E_{0}}{E_{a}}=\frac{\hbar ^{4}E_{0}}{m_{e}^{2}e^{5}}
\end{equation}
becomes equal to unity for a laser radiation intensity equal to $%
m_{e}^{4}e^{10}c/4\pi \hbar ^{8}\simeq 10^{16}$ W/cm$^{2}$. For small but
finite values of this parameter, i.e. in the limit $I<10^{16}$ W/cm$^{2}$,
the atom is not ionized, unless the multiphoton processes come into play,
the EMW--matter interaction can be described within the framework of
perturbation theory. When the parameter $E_{0}/E_{a}$ approaches unity, the
potential inside the atom changes its form and the so-called tunnel
ionization becomes possible. The tunnel ionization probability is given by
the Keldysh formula \cite{Keldysh} 
\begin{equation}
w=\omega _{a}\exp \left[ {-\frac{2W_{b}}{\hbar \omega _{0}}f(\gamma _{K})}%
\right] ,  \label{eq:tunion}
\end{equation}
where the function $f(\gamma _{K})\approx 2\gamma _{K}/3$ for $\gamma
_{K}\ll 1$ and $f(\gamma _{K})\approx \ln 2\gamma _{K}-1/2$ for $\gamma
_{K}\gg 1$. The adiabatic parameter $\gamma _{K}$ is defined as 
\begin{equation}
\gamma _{K}=\omega _{0}\frac{\sqrt{2m_{e}W_{b}}}{eE_{0}}=\sqrt{\frac{2\hbar
\omega _{a}}{a_{0}^{2}m_{e}c^{2}}}.
\end{equation}
Here introduced is the EMW dimensionless amplitude,

\begin{equation}
a_{0}=\frac{eE_{0}}{m_{e}\omega _{0}c}.  \label{eq:dimla}
\end{equation}

In the limit $\gamma _{K}\ll 1$, i.e. for a relatively strong
electromagnetic wave, Eq. (\ref{eq:tunion}) corresponds to the ionization
probability by a constant electric field, 
\begin{equation}
w=2\omega _{a}\frac{E_{a}}{E_{0}}\exp \left( {-}\frac{{2E_{a}}}{{3E_{0}}}%
\right) .
\end{equation}
For intensities larger than $m_{e}^{4}e^{10}c/4\pi \hbar ^{8}\simeq 10^{16}$
W/cm$^{2}$ the deformation of the potential inside the atoms caused by the
laser pulse field becomes so strong that the electron energy level becomes
larger than the maximum value of the potential. As a result, the electron
appears as if in a free state and leaves the atom. Due to the periodicity of
the electric field, there is a probability that the electron will return
after a half of the wave period. Recollisions with the ions lead to the
generation of high order harmonics \cite{Corkum}. However, for a very strong
electromagnetic wave the effects of the wave magnetic field decrease this
probability. In this case the matter becomes ionized in one optical period
and plasma processes start to play a key role.

\begin{figure}[tbp]
\includegraphics[width=9cm]{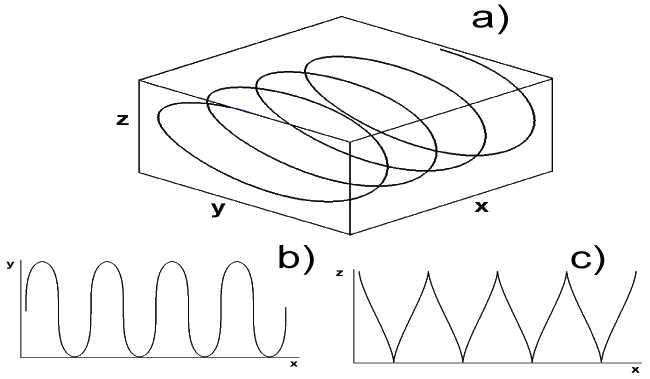}
\caption{Projectories of the charged particle trajectory, when it interacts
with the elliptically polarized EMW.}
\label{fig:01}
\end{figure}

Under the action of the electromagnetic wave the plasma electrons oscillate
at the wave frequency. In the limit $v<<c$ their quiver velocity is
approximately equal to $v_{E}=eE_{0}/m_{e}\omega _{0}$. In the
non-relativistic limit, when $v_{E}/c\ll 1$ or $a_{0}\ll 1$, the electron
quiver amplitude is smaller than the laser wavelength, $\lambda _{0}$. Under
the action of the electromagnetic wave, given by the vector potential $%
A_{\bot }(x-ct)$, the electrons oscillate at the wave frequency. From the
equations of the motion we obtain that the transverse component of the
generalized momentum $p_{\bot }-eA_{\bot }(x-ct)/c$ is constant. The
particle energy and the longitudinal momentum component are related as \cite%
{LLTF} 
\begin{equation}
\sqrt{m_{e}^{2}c^{4}+p_{\bot }^{2}+p_{\parallel }^{2}}-p_{||}c=h.
\end{equation}
In the reference frame where the particle was at rest before interaction
with the laser pulse, the particle kinetic energy 
\begin{equation}
K=m_{e}c^{2}\left( {\sqrt{1+(p/m_{e}c)^{2}}-1}\right)
\end{equation}
and momentum $\mathbf{p=(}p_{||},p_{\bot }\mathbf{)}$ are given by
expressions $K=m_{e}c^{2}\left| {a_{\bot }}\right| ^{2}/2$, $p_{\bot
}=m_{e}c\,a_{\bot }$, $p_{||}=m_{e}c\left| {a_{\bot }}\right| ^{2}/2$. Here $%
a_{\bot }=eA_{\bot }(x-ct)/m_{e}c^{2}$. For $|a_{\bot }|>1$ the particle
acquires a relativistic energy, and the longitudinal component of its
momentum is larger than the transverse component. Fig. \ref{fig:01} shows a
typical trajectory of the charged particle in the electromagnetic wave.

The EMW behavior in a plasma differs from its behavior in vacuum, and
depends on the electron density. In a plasma with a density $n$, a
displacement of the electrons with respect to the ions generates the
electric field. Its ratio to the laser electric field is $E/E_{0}=4\pi
ne^{2}/m_{e}\omega _{0}^{2}=(\omega _{pe}/\omega _{0})^{2}=n/n{{{{_{cr}}}}}$%
, where $\omega _{pe}=\sqrt{4\pi ne^{2}/m_{e}}$ is the Langmuir frequency
and $n_{cr}={m_{e}\omega _{0}^{2}{{/{4\pi e^{2}}}}}$ is the critical
density. The dimensionless parameter
\begin{equation}
\frac{\omega _{pe}}{\omega _{0}}=\sqrt{\frac{n}{n_{cr}}}=\sqrt{\frac{4\pi
ne^{2}}{m_{e}\omega _{0}^{2}}}
\end{equation}
is a measure of the plasma collective response to a periodic electromagnetic
field.

When an EMW propagates through a plasma, its group velocity, $v_{g}=\partial
\omega /\partial k$, and phase velocity, $v_{ph}=\omega /k$, are not equal
to each other and are related as $v_{g}v_{ph}=c^{2}$. While in vacuum the
dispersion equation for the frequency, $\omega $, and wave vector,$k$, takes
the form $\omega ^{2}=k^{2}c^{2}$, in a plasma it becomes $\omega
^{2}=k^{2}c^{2}+\omega _{pe}^{2}$. This dispersion equation can be rewritten
as $k=\sqrt{\omega ^{2}-\omega _{pe}^{2}}/c$, which shows that an EMW with a
frequency below the Langmuir frequency cannot propagate through the plasma
and that the electromagnetic field evanescence length in a high density
plasma is of the order of the collisionless skin depth, $d_{e}=c/\omega
_{pe} $, i.e. an overdense plasma with the electron density higher than the
critical density is not transparent.

The collective response of the plasma, in addition to the transverse
electromagnetic mode, exhibits longitudinal plasma oscillations, i.e.,
Langmuir waves. The electric field in a Langmuir wave oscillates with
frequency $\omega =\omega _{pe}$. The group velocity of the Langmuir waves
vanishes, $v_{g}=\partial \omega _{pe}/\partial k=0$, and their phase
velocity is determined by the wave number.

Relativistic effects change the dispersion equation due to the dependence of
the Langmuir frequency on the wave amplitude. As found in Ref. \cite{A-P},
the frequency of a longitudinal wave depends on its amplitude $%
a_{L}=eE/m_{e}\omega _{pe}c$ as $\omega \approx \omega _{pe}(1-3a_{L}^{2}/4)$
for $a_{L}\ll 1$ and as $\omega \approx \omega _{pe}/\sqrt{8a_{L}}$ in the
case $a_{L}\gg 1$.

For a circularly polarized electromagnetic wave the dispersion equation
takes the form: 
\begin{equation}
\omega ^{2}=k^{2}c^{2}+\frac{\omega _{pe}^{2}}{\sqrt{1+a_{0}^{2}}}.
\end{equation}
We see that the effective critical density increases as the EMW amplitude
grows, i.e., the plasma is more transparent to high intensity
electromagnetic radiation.

Large amplitude, finite length pulses of electromagnetic and Langmuir waves
do not propagate independently since they are coupled by nonlinear
processes. The Langmuir wave that is generated by an ultra short laser
pulse, being left behind in the plasma and thus called the wake wave. It is
of special interest since the structure of the electric field of this wake
wave is favorable for charged particle acceleration. In a low density plasma
the phase velocity of the wake wave can be very close to the speed of light
in vacuum. In analogy to linear accelerators that use electric fields in the
radio-frequency range in Ref. \cite{T-D} it was proposed to use the wake
field for charged particle acceleration.

The dimensionless amplitude, Eq. (\ref{eq:dimla}), is equal to the electron
quiver momentum normalized to $m_{e}c$. For a pulse with an intensity
corresponding to $a_{0}>1$, relativistic effects must be taken into account.
The intensity of a linearly polarized electromagnetic wave can be written
via $a_{0}$ as
\begin{equation}
I_{L}=\frac{\pi }{2}\frac{a_{0}^{2}}{\lambda _{0}^{2}}\frac{{m_{e}c^{3}}}{%
r_{e}}\approx 1.37\times 10^{18}\times a_{0}^{2}\times \left( {\frac{1\mu m}{%
\lambda _{0}}}\right) ^{2}\frac{W}{cm^{2}}.
\end{equation}

If the wave is focused into a one wavelength spot, this intensity
corresponds to the power $\mathcal{P}=a_{0}^{2}\times 43\;$GW. At present
laser intensities have reached a level above $10^{22}$W/cm$^{2}$ \cite{YAN}.

When the electron energy approaches $3m_{e}c^{2}$, electron-positron pairs
are generated during electron-nuclei collisions, \cite{QED}. The cross
section of this process is given by
\begin{equation}
\sigma _{\pm }=\frac{28}{27\pi }r_{e}^{2}(\alpha Z)^{2}\left[ {\ln \left( 
\sqrt{1+\left( \frac{p}{m_{e}c}\right) ^{2}}\right) }\right] .
\end{equation}

Here $Ze$ is the nucleus electric charge and $\alpha =e^{2}/\hbar c=1/137$
is the fine-structure constant. Positron generation in a plasma has been
discussed in a number of publications (e.g. see \cite{BKZS}) and was
observed in the terawatt laser plasma interaction experiments \cite{POS}.
We note a discussion of the pion and muon production in electron-positron
and gamma plasmas \cite{IKUZ}. 

\subsection{Interaction of EMW with Plasmas in the Radiation-Dominated Regime}

The dimensionless parameters characterizing the electromagnetic emission by
an electron are the ratio between the classical electron radius and the
electromagnetic wavelength, $r_{e}/\lambda _{0}=e^{2}\omega _{0}/2\pi
m_{e}c^{3}$, and the ratio between the photon energy and the electron rest
mass energy, $\hbar \omega _{0}/m_{e}c^{2}$.

When an electron moves under the action of the electric and magnetic field
of a wave, it emits electromagnetic radiation. The intensity of this
radiation is given by the formula $W=(2e^{2}/3m_{e}^{2}c^{3})(dp_{\mu
}/d\tau )^{2}$, where $p_{\mu }$ is the particle 4-momentum and $\tau $ is
its proper time. When an ultrarelativistic charged particle moves along a
circular trajectory in a circularly polarized electromagnetic wave, the
radiation intensity is $W=\left( {4\pi r_{e}/3\lambda _{0}}\right) \omega
_{0}\,m_{e}c^{2}a_{0}^{4}$. We see that the relative role of the radiation
damping force is determined by the dimensionless parameter$\varepsilon
_{rad} $, which is equal to 
\begin{equation}
\varepsilon _{rad}=\frac{4\pi r_{e}}{3\lambda _{0}}.
\end{equation}%
By comparing the energy radiated by the particle per unit of time with the
maximum energy gain in the electromagnetic wave $\partial _{t}\mathcal{E}%
=\omega _{0}m_{e}c^{2}a_{0}$, we obtain that the radiation effects become
dominant at $a_{0}\geq a_{rad}=\varepsilon _{rad}^{-1/3}$ , i. e. in the
limit $I>10^{23}W/cm^{2}$ for 1 $\mu $m wavelength laser \cite{BEKT,RAD}. In
the limit of a relatively low amplitude laser pulse, $a_{0}\ll a_{rad}$,~the
momentum of an electron moving in a circularly polarized electromagnetic
wave in a plasma scales with the laser pulse amplitude as $p=m_{e}ca_{0}$,
while in the limit $a_{0}\gg a_{rad}$, it scales as $p=m_{e}c(a_{0}/%
\varepsilon _{rad})^{1/4}$.

\begin{figure}[tbp]
\includegraphics[width=9cm]{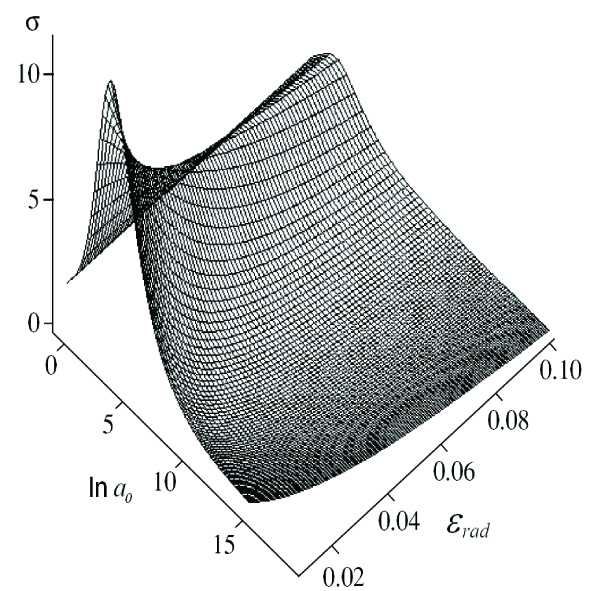} 
\caption{Scattering cross section dependence on the EMW amplitude and
wavelength.}
\label{fig:02}
\end{figure}

Quantum effects become important, when the energy of the photon generated by
Compton scattering is of the order of the electron energy, i.e. $\hbar
\omega _{m}\approx \mathcal{E}_{e}$. An electron with energy $\mathcal{E}%
_{e}=\gamma m_{e}c^{2}$ rotates with frequency $\omega _{0}$ in a circularly
polarized wave propagating in a plasma and emits photons with frequency $%
\omega _{m}=\gamma ^{3}\omega _{0}$. We obtain that quantum effects come
into play when $\gamma \geq \gamma _{Q}=\sqrt{m_{e}c^{2}/\hbar \omega _{0}}$%
. For an electron interacting with one-micron laser light we find $\gamma
_{Q}\approx 600$. From the previous analysis we obtain that the quantum
limit on the electron gamma factor corresponds to 
\begin{equation}
a_{Q}=\frac{2e^{2}m_{e}c}{3\hbar ^{2}\omega _{0}}.
\end{equation}
The energy flux reemitted by the electron is equal to $e(E\cdot
v)=\varepsilon _{rad}\omega _{0}\gamma ^{2}p_{\bot }^{2}/m_{e}$. The total
scattering cross section defined as the ratio of the reemitted energy to the
Poynting vector $P=cE_{0}^{2}/4\pi $, is given by 
\begin{equation}
\sigma =\sigma _{T}\frac{\gamma ^{2}}{1+\varepsilon _{rad}^{2}\gamma ^{6}},
\end{equation}
where the Thomson scattering cross section is $\sigma _{T}=8\pi
r_{e}^{2}/3=6.65\times 10^{-25}cm^{2}$. We see that, as the wave amplitude
increases in the range $1\ll a_{0}\ll a_{rad}$, the scattering cross section
increases according to the law $\sigma =\sigma _{T}(1+a_{0}^{2})$ and
reaches its maximum $\sigma =\sigma _{T}a_{rad}^{2}$ at $a_{0}\approx
a_{rad} $; for $a_{0}\gg a_{rad}$, it decreases according to the law $\sigma
=\sigma _{T}a_{rad}^{3}/a_{0}$. In Fig. \ref{fig:02} we show the scattering
cross section dependence on the EMW amplitude and wavelength. 

In the radiation-dominated regime of the EMW interaction with charged
particles, i.e. at $a_{0}>a_{rad}$, the emitted gamma quanta can produce
secondary electron-positron pairs, which in turn emit gamma ray photons,
producing an avalanche of $\gamma $ rays and electron-positron pairs \cite%
{Bell}.

\subsection{Probing Nonlinear Vacuum}

When the amplitude of the electromagnetic wave approaches the critical
electric field of quantum electrodynamics (also called the ``Schwinger
field''), vacuum becomes polarized and electron-positron pairs are created
in vacuum \cite{QED,DGi}. On a distance equal to the Compton length, $%
\lambda _{C}=\hbar /m_{e}c$, the work of the critical field on an electron
is equal to the electron rest mass energy, $m_{e}c^{2}$, i.e. $%
eE_{QED}\lambda _{C}=m_{e}c^{2}$. The dimensionless parameter

\begin{equation}
\frac{E}{E_{QED}}=\frac{e\hbar E}{m_{e}^{2}c^{3}}
\end{equation}
becomes equal to unity for an electromagnetic wave intensity of the order of

\begin{equation}
I=\frac{c}{r_{e}\lambda _{C}^{2}}\frac{m_{e}c^{2}}{4\pi }\approx 4.7\times
10^{29}\frac{W}{cm^{2}}.
\end{equation}

For such ultrahigh intensities the effects of nonlinear quantum
electrodynamics play a key role: an electromagnetic wave excites virtual
electron-positron pairs. An observable manifestation of this process could
be detection of light birefringence during the propagation of an
electromagnetic wave in a strong electric or magnetic field in vacuum. The
cross section for the photon-photon interaction in the limit $\hbar \omega
\ll m_{e}c^{2}$ is given by
\begin{equation}
\sigma _{\gamma \gamma \to \gamma \gamma }=\frac{973}{10125}\frac{\alpha ^{2}%
}{\pi ^{2}}r_{e}^{2}\left( {\frac{\hbar \omega }{m_{e}c^{2}}}\right) ^{6},
\end{equation}
where $\hbar \omega $ is the photon energy (see \cite{QED}). This cross
section reaches its maximum, $\sigma _{\max }\approx 10^{-20}cm^{2}$, for $%
\hbar \omega \approx m_{e}c^{2}$, i.e. for the interactions of photons in
the gamma range. Also attention is focused on the process of
electron-positron pair creation in vacuum by an electromagnetic wave. For an
electric field small compared to $E_{QED}$, this process is sub-barrier,
similarly to the tunnel ionization of atom discussed above [see Eq. (\ref%
{eq:tunion})]. The probability of electron-positron pair creation per unit
volume and per unit time is exponentially small and is given by
\begin{equation}
w=\left( \frac{\alpha c}{\pi ^{2}\lambda _{C}^{4}}\right) \left( \frac{E}{%
E_{QED}}\right) ^{2}exp\left( -\pi \frac{E_{QED}}{E}\right) .  \label{eq:w}
\end{equation}
Here $\lambda _{C}=\hbar /m_{e}c$ is the Compton length and $\alpha
=e^{2}/\hbar c=1/137$ is the fine structure constant.

We may formally estimate the number of electron-positron pairs produced by a
10 fs long laser pulse in a volume $V=\lambda ^{3}=10^{-12}$cm$^{3}$ as $%
N_{\pm }=wV\tau _{p}$. It is easy to show that $N_{\pm }$ is equal to one
pair for a laser intensity equal to $I=10^{26}$W/cm$^{2}$ (a more detailed
description of this process can be found in \cite{NAR} and in Refs. \cite%
{QED,VSP,NBMP}). Obviously, this latter number is overestimated because the
minimum needed energy is by many orders of magnitude larger than the total
energy of the laser pulse. At intensities of the order $I=10^{30}$W/cm$^{2}$
Eq. (\ref{eq:w}) is not applicable and a depletion of the laser pulse must
be taken into account. The electromagnetic pulse depletion due to its energy
conversion into electron-positron pairs has been studied in Ref. \cite{BFP}. 
\begin{figure}[tbp]
\includegraphics[width=9cm]{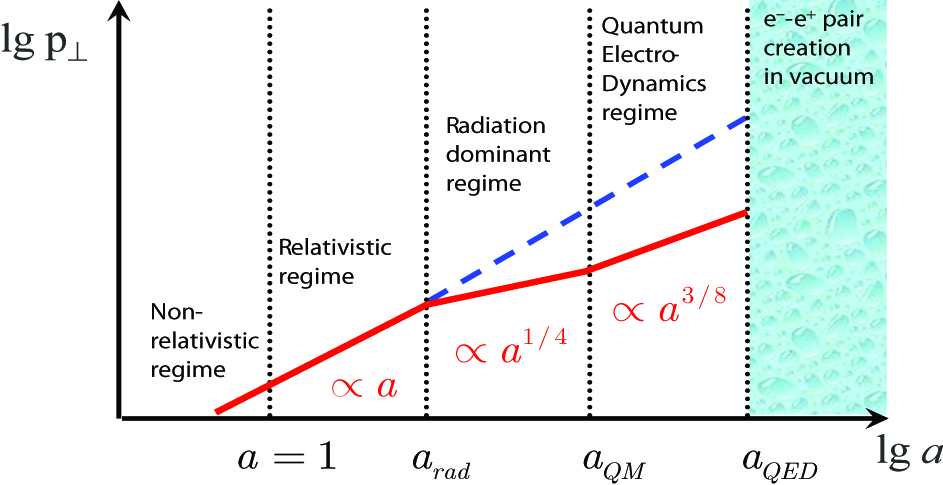} 
\caption{Various regimes of relativistically strong EMW interaction with
plasmas.}
\label{fig:03}
\end{figure}

The nonlinear dependence of the vacuum susceptibilities on the
electromagnetic-field amplitude results in the finite value of the Kerr
constant of vacuum. It can be found to be 
\begin{equation}
K_{K}=\frac{7\alpha }{90\pi }\frac{\lambda _{C}^{3}}{m_{e}c\lambda _{0}}
\end{equation}
The Kerr constant in vacuum for $\lambda _{0}=1\mu m$ is of the order of $%
10^{27}cm^{2}/erg$, which is a factor $10^{20}$ smaller than for water. As
shown in Ref. \cite{Rozanov}, in a QED nonlinear vacuum two
counterpropagating electromagnetic waves mutually focus each other. A
nonlinear modification of the refraction index in vacuum within the
framework of the Heisenberg-Euler approximation is characterized by the
critical value of the electromagnetic wave power 
\begin{equation}
\mathcal{P}_{QED}=45\pi ^{2}\frac{cE_{QED}^{2}\lambda _{0}^{2}}{4\pi \alpha }.
\end{equation}
When the electromagnetic wave power exceeds this value, the cross modulation
nonlinear effects affect the wave propagation. We see that the critical
power, $\mathcal{P}_{cr}$, depends only on the laser pulse wavelength, $%
\lambda _{0}$, and on fundamental constants. It is easy to show that for $%
\lambda _{0}=1\mu $m the critical power $\mathcal{P}_{cr}=cE^{2}w^{2}/4$,
where $w$ is the laser beam waist. For the mutual self-focusing $\mathcal{P}%
_{cr}=2.5\times 10^{24}$W can be found to be for $\lambda _{0}=1\mu m$.

Nonlinear modifications of the vacuum refraction index lead to the vacuum
birefringence \cite{Rozanov}, to the four-wave interaction \cite{four-wave}, 
to the high order harmonic generation \cite{Narozhny-Fedotov}, 
and to the laser-photon splitting and merging
\cite{DIP} (see also
review articles \cite{MTB-06}, \cite{MSh-06,SHHK-06}). According to
Ref. \cite{Unruh} the Unruh radiation intensity of the electron moving in
the field of a strong electromagnetic wave becomes comparable with the
nonlinear Thomson scattering intensity under the condition $4\pi a_{0}\hbar
\omega _{0}/m_{e}c^{2}\approx 1$. The multi-photon Compton scattering during
the collision of counter-propagating laser beams and ultrarelativistic
electron bunches leading to the gamma quanta generation 
\begin{equation}
e^{-}+n\hbar \omega _{0}\rightarrow \hbar \omega _{\gamma },
\end{equation}
with their subsequent interaction with the laser light accompanied by the
electron -positron pair creation in vacuum via the Breit-Wheeler process 
\begin{equation}
\hbar \omega _{\gamma }+n\hbar \omega _{0}\rightarrow e^{-}+e^{+}
\end{equation}
has been investigated in Ref. \cite{Burke}. In Ref. \cite{ErL} the cross
section of the Breit-Wheeler process on the laser pulse intensity has been
investigated.

Various regimes of the relativistically strong EMW interaction with plasmas
are illustrated in Fig. \ref{fig:03}.

\subsection{EMW Parameters under Space Plasma Conditions}

\begin{figure}[tbp]
\includegraphics[width=9cm]{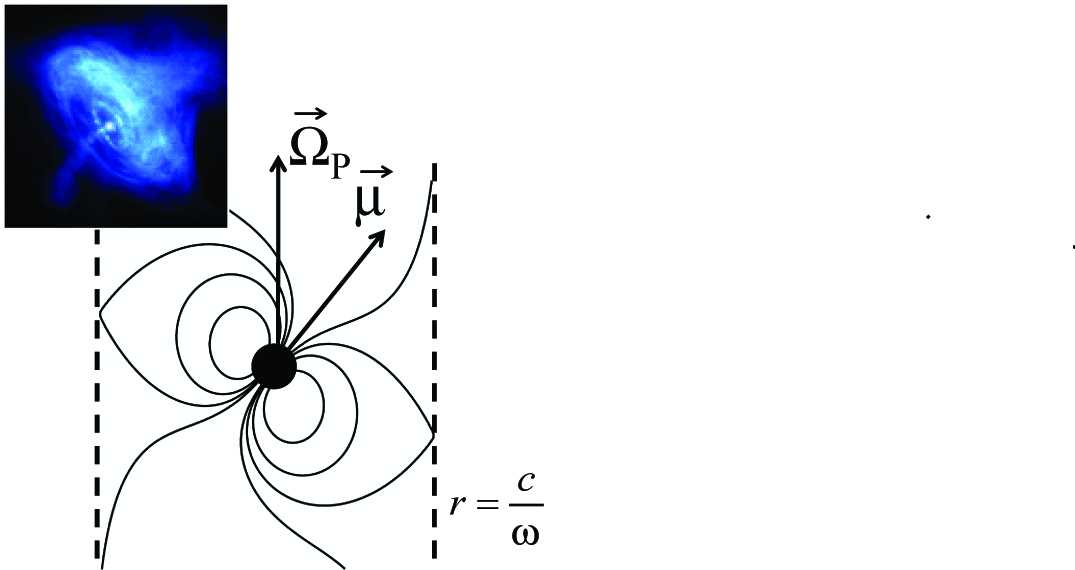} 
\caption{Pulsar magnetosphere. The inset: The Crab pulsar \protect\cite%
{Hester}.}
\label{fig:04}
\end{figure}

In one of the first works on the charged particle acceleration by strong
electromagnetic waves in astrophysical plasmas, pulsars \cite{BG} have been
considered as sources of ultraintense radiation \cite{GO}. Pulsars are
considered to be oblique rotators with non-parallel rotation and magnetic
dipole axes, as illustrated in Fig. \ref{fig:03}. The power of
magneto-dipole radiation is given by the expression $\ $%
\begin{equation}
W=\frac{2\mu ^{2}\sin ^{2}\chi \Omega _{P}^{4}}{3c^{3}},
\end{equation}
where $\mu $ is the magnetic momentum, $\chi $ is an angle between the
rotational and magnetic dipole axes, and $\Omega _{P}$ is the pulsar
rotation frequency. Even for parallel magnetic and angular moments, i.e. for 
$\chi =0$, the expression $W=(2/3)\mu ^{2}\Omega _{P}^{4}/c^{3}$ gives the
pulsar electromagnetic energy losses, as it follows from the theoretical
model of the pulsar magnetosphere \cite{GR}. The magnetic moment is related
to the pulsar magnetic field and radius as $\mu \approx Br_{P}^{3}$. For
typical values of $r_{P}=10^{6}$cm and $B=10^{12}\ $G we obtain $\mu
=10^{30}\ $G\ cm$^{3}$. The electromagnetic wave intensity at the distance $%
r $ is equal to $I=W/4\pi r^{2}$. At the wave zone boundary, $r=c/\Omega
_{P} $, the dimensionless amplitude of the electric field is 
\begin{equation}
a_{P}=\frac{e\mu \Omega _{P}^{2}}{m_{e}c^{4}}.
\end{equation}
For the Crab pulsar with the rotation frequency $\Omega _{P}=200\ $s$^{-1}$
we find $a_{P}=2\times 10^{10}$.

According to Ref. \cite{BEKT,RAD} in the limit of high radiation intensity
the effects of the radiation damping should be incorporated into the theory
of the electromagnetic wave interaction with plasmas. A dimensionless
parameter, 
\begin{equation}
\varepsilon _{rad}=\frac{2e^{2}\Omega _{P}}{3m_{e}c^{3}},
\end{equation}
gives a value of the wave amplitude, $a_{rad}=\varepsilon _{rad}^{-1/3}$,
above which the radiation damping cannot be neglected. For $\Omega _{P}=200\ 
$s$^{-1}$ this yields $a_{rad}=10^{7}$, which is substantially less than
above found value of $a_{P}=2\times 10^{10}$.

In the case of laser - plasma interaction for a typical laser wavelength of $%
1\mu $m the dimensionless amplitude $a_{rad}$ corresponds to an intensity of
the order of $10^{23}$W/cm$^{2}$ which can be achieved by tight focusing of
the PW power laser beams onto the one-lambda size focus spot. We see that
the laser plasmas can be used for modeling the radiation damping effects,
which are important for relativistic astrophysics.

\section{Acceleration of Charged Particles in the EMW Interaction with
Plasmas}

\label{LACP} General requirements for the laser accelerator parameters are
principally the same as for standard accelerators of charged particles \cite%
{ACC-EM}, i. e. they should have a reasonable acceleration scale length, a
high enough efficiency and the required maximal energy, a high quality,
emittance and luminosity of charged particle beams. In the 1940-s Enrico
Fermi paid attention to the high energy limit of $\approx 1\ $PeV$=10^{15}\ $%
eV for accelerated particles, which could be reached under terrestrial
conditions, when the accelerator size is limited by the equator
circumference. These limitations resulted in the 1950-ties in the proposal
to use collective electric fields excited in a plasma (collective methods of
acceleration) in order to accelerate charged particles \cite{VEK}.

\subsection{Electron Accelerator}

\label{EA}

Wakefield acceleration has been proposed in Ref. \cite{CTT} for the
generation of ultra high energy cosmic rays. Below we describe the wake
field acceleration mechanism using as an example the LWFA scheme. 
\begin{figure}[tbp]
\includegraphics[width=9cm]{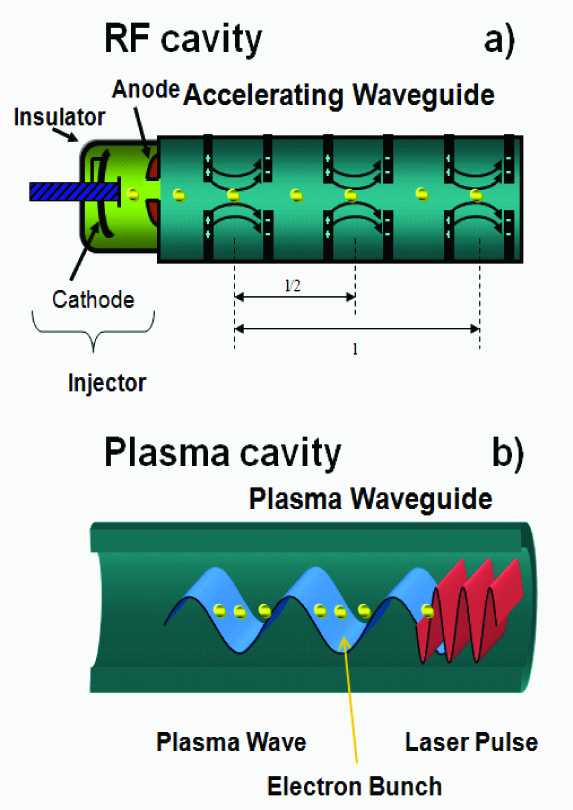} 
\caption{Schematic view of the standard linear accelerator of charged
particles, a) and the LWFA, b).}
\label{fig:05}
\end{figure}

Under the condition of minimum laser energy the one stage LWFA accelerator
scaling is described as it follows \cite{T-D}. The electric field in a
plasma has the form of a wave propagating with a phase velocity, $v_{ph,W}$.
A gamma factor corresponding to the wave phase velocity is given by the
expression $\gamma _{ph,W}=(1-v_{ph,W}^{2}/c^{2})^{-1/2}$. A condition of
the wake wave synchronization with the driver laser pulse yields $%
v_{ph,W}=v_{g,las}$, where $v_{g,las}\approx c(1-\omega _{pe}^{2}/2\omega
_{0}^{2})$ is the laser pulse group velocity. The wavelength of the weakly
nonlinear wake wave is $\lambda _{p}=\lambda _{0}\gamma _{ph,W}$. Assuming
the electrostatic potential in the wake is equal to $m_{e}c^{2}/e$, we
obtain for the fast electron gamma factor $\gamma _{e}=2\gamma _{ph,W}^{2}$.
The acceleration length is given by $l_{acc}=\lambda _{p}\gamma _{ph,W}^{2}$%
, i.e. $l_{acc}=\lambda _{0}\gamma _{ph,W}^{3}$. This gives a relationship
between the acceleration length and the fast electron energy: $%
l_{acc}=\lambda _{0}\gamma _{e}^{3/2}$. For $\lambda _{0}=1\ \mu $m and $%
\gamma _{e}=10^{6}$ we obtain $l_{acc}\approx 1$km \cite{KanTeV}.

\begin{figure}[tbp]
\includegraphics[width=9cm]{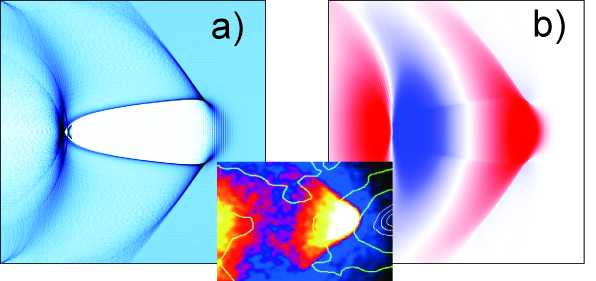} 
\caption{2D PIC simulations show that the electrons pushed away by the
ponderomotive pressure of the laser pulse form the "bow wave" \protect\cite%
{NEW}. The electron density distribution (a) clearly shows the
'swallow-tail' formation during the wake wave breaking in the first period
of the wave behind the laser pulse. The wakefield (the x-component of the
electric field) is excited by the laser pulse in an underdense plasma (b).
Inset: The bow wave formed by colliding galaxies in the Bullet Cluster 
\protect\cite{DClowe}.}
\label{fig:06}
\end{figure}

In the opposite limit, when the laser transverse width $r_{las}\le \lambda
_{p}$, we need to take into account the formation of an electron density
cavity moving with the group velocity of the laser pulse (see Fig.\ref%
{fig:06}, where the wake wave left behind the ultra short laser pulse in the
underdense plasma is shown). The cavity's transverse size is determined by
the laser pulse width and its length is of the order of the Langmuir wave
wavelength. In this limit, the wavelength depends on the amplitude of the
Langmuir wave, which in turn depends on the laser pulse intensity. For a
given laser pulse width the electrostatic potential in the cavity is of the
order of $\phi \approx \pi n_{0}er_{las}^{2}$, and the group velocity of a
narrow laser pulse is determined by its width, i.e. $\gamma _{ph,W}\approx
r_{las}/\lambda _{0}$. As a result we find the electron energy scaling \cite%
{NEW}: 
\begin{equation}
\gamma _{e}=\frac{r_{las}^{4}}{\lambda _{0}^{2}\lambda _{p}^{2}}.
\end{equation}
It does not depend on the laser pulse amplitude provided $a_{0}>e\phi
/m_{e}c^{2}$. The laser energy depletion length in this limit is given by 
\begin{equation}
l_{dep}=a_{0}l_{las}\left( \frac{\lambda _{p}}{\lambda _{0}}\right) ^{2},
\end{equation}
i. e. it is by a factor $a_{0}$ greater than in the 1D case.

Considering the laser electron accelerator for the applications in the high
energy physics, we find that its parameters should satisfy several
conditions in addition to the requirement on the maximum particle energy.
Parameters of fundamental importance such as the luminosity characterize the
number of reactions produced by the particles in colliding beams of a
collider. The luminosity is given by the expression 
\begin{equation}
\mathcal{L}=f\frac{N_{1}N_{2}}{4\pi \sigma _{y}\sigma _{z}},
\end{equation}
where $N_{1}$ and $N_{2}$ are the numbers of particles in each of the beams, 
$\sigma _{y}$ and $\sigma _{z}$ are the transverse size of the beam in the $%
y $ and $z$ directions, and $f$ is the frequency of the beam collisions. A
product of the luminosity and the reaction cross section gives the reaction
rate. We see that the luminosity can be increased by increasing the particle
number in a bunch, $N_{j}$, and/or by increasing the repetition rate, $f$,
or by decreasing the transverse size of the bunch, $\sigma _{i}$, by
focusing the particle beam into the minimum size focal spot. The focal spot
size depends on the beam emittance, which is defined as the surface occupied
by the bunch in the phase plane ($(y,p_{y})$ or $(z,p_{z})$). A calculation
under the assumption of a round transverse shape of the beam ($\sigma
_{y}=\sigma _{z}=r$) results in the expression given by the integral 
\begin{equation}
\varepsilon _{\bot }=\frac{1}{\pi }\int drdr^{\prime },
\end{equation}
where $r$ is the transverse size of the bunch and $r^{\prime }=dr/dx=dr/cdt$ 
\cite{ACC-EM}.

The transverse dynamics of the electron in the field of the wake wave is
described by the equation (see for example Ref. \cite{B-PoP}) 
\begin{equation}
\frac{d}{dt}\left( \gamma _{||}\frac{dr}{dt}\right) +\omega _{pe}^{2}r=0,
\end{equation}
where the electron gamma factor depends on time as $\gamma _{||}(t)=\gamma
_{e}(1-t^{2}/t_{acc}^{2})$ with $\gamma _{e}=\gamma _{ph,W}^{2}$ and $\gamma
_{ph,W}=\omega _{0}/\omega _{pe}$. In the limit $\gamma _{||}\gg 1$ the
electron transverse oscillations are described by the dependence of the
radial displacement on time: 
\begin{equation}
r(t)=r_{inj}\left( \frac{\gamma _{inj}}{\gamma _{||}(t)}\right) ^{1/4}\cos %
\left[ \int_{t_{inj}}^{t}\omega _{b}(t^{\prime })dt^{\prime }\right] ,
\end{equation}
where $\omega _{b}(t)=\omega _{pe}\gamma _{||}(t)^{2}$ is the betatron
oscillation frequency and $r_{inj}$ and $\gamma _{inj}\approx \gamma _{ph,W}$
are the radial coordinate and the electron energy at the injection time, $%
t_{inj}$ normalized on $m_{e}c^{2}$. Calculating the transverse emittance,
we find $\varepsilon _{\perp }=\pi \kappa ^{2}(\omega _{pe}/\omega _{0})^{3}$
mm\ mrad, with $\kappa =r_{inj}/\lambda _{p}$. The normalized emittance, $%
\varepsilon _{N}=\varepsilon _{\perp }\gamma _{e}$, is equal to $\varepsilon
_{\bot }=\pi \kappa ^{2}(\omega _{pe}/\omega _{0})$~mm~mrad.

The electron motion in the electric field of the wake plasma wave is
characterized by the structure of the phase plane ($p_{x},X=x-v_{ph}t$). A
calculation of the energy spectrum of fast electrons is done in Refs. \cite%
{B-PoP,QME}). It uses the property of electrons injected at the breaking
point to move along the separatrix. The electrons, whose trajectories lie on
the separatrix, where they are uniformly distributed, near the top of the
separatrix have an electron momentum dependence on the coordinate 
\begin{equation}
p_{x}=p_{m}(1-X^{2}\omega _{pe}^{2}/c^{2}a_{0})=p_{m}(1-t^{2}/t_{acc}^{2}).
\end{equation}
The distribution function of the electrons at the target has the form 
\begin{equation}
f(t,\mathcal{E})=(n_{b}\omega _{pe}/\sqrt{2}ca_{0})\delta (\mathcal{E}-%
\mathcal{E}_{m}(1-t^{2}/t_{acc}^{2})).
\end{equation}
Here $\delta (z)$ is the Dirac delta function and we have assumed that the
electrons are ultrarelativistic with $\mathcal{E}=p_{x}c$ and $\mathcal{E}%
_{m}=p_{m}c$ . In order to find the energy spectrum of the electrons on the
target, we must integrate the function on time in the limits between $%
-t_{acc}$ and $t_{acc}$. We obtain

\begin{equation}
\frac{d\mathcal{N}(\mathcal{E})}{d\mathcal{E}}=\frac{n_{b}\omega _{pe}}{%
\sqrt{2}ca_{0}}\int\limits_{-t_{acc}}^{t_{acc}}\delta \left( \mathcal{E}-%
\mathcal{E}_{m}\left( 1-\frac{t^{2}}{t_{acc}^{2}}\right) \right) dt=\frac{%
n_{b}\omega _{pe}}{2\sqrt{2}ca_{0}\sqrt{\mathcal{E}_{m}(\mathcal{E}_{m}-%
\mathcal{E})}},
\end{equation}%
i.e. the particle spectrum has a typical form $\propto 1/\sqrt{\mathcal{E}%
_{m}-\mathcal{E}}$ near maximum energy.

Using the above given relationships and estimating a maximum number of
particles in a bunch as $N\approx \kappa ^{2}n_{e}\lambda _{p}^{3}$, we
obtain the luminosity to be equal to 
\begin{equation}
\mathcal{L}=10^{34}\left( \frac{f}{10KHz}\right) \left( \frac{\kappa }{0.1}%
\frac{\lambda _{0}}{r_{inj}}\right) ^{2}\left( \frac{\gamma _{e}}{10^{6}}%
\right) ^{3/2}\ \frac{1}{\mathrm{cm}^{2}\mathrm{s}}.
\end{equation}
Here we assume a round transverse shape of the bunch with $r\approx
r_{inj}(\gamma _{inj}/\gamma _{e})^{1/4}$. Utilization of flat bunches with $%
\sigma _{y}\gg \sigma _{z}$ allows and to achieve larger luminosity \cite%
{kando-07}. In addition, in the case of flat beams the space charge effects
and beamsstahlung can be weakened. We notice that the radiation damping
effects on the LWFA operation have been considered in Ref. \cite{MSSEL}.

\subsection{Ion Accelerator}

\label{IA}

The mechanism of laser acceleration of ions (protons and other ions) is
determined by the electric field set up by the space charge separation of
hot or energetic electrons and the ions. The exact mechanisms entering into
the energy transfer from the fast electron to the ion energy depends on the
specific conditions of the laser-target interaction (see review articles 
\cite{MTB-06} and \cite{BFB}). The proton generation is a direct consequence
of the electron acceleration.

The typical energy spectrum of laser accelerated particles observed both in
experiments and in computer simulations can be approximated by a
quasi-thermal distribution with a cut-off at a maximum energy. On the other
hand, the applications require high quality proton beams, i.e. beams with
sufficiently small energy spread $\Delta \mathcal{E}_{i}/\mathcal{E}_{i}\ll
1_{i}$. For example, for hadron therapy it is highly desirable to have a
proton beam with $\Delta \mathcal{E}_{i}/\mathcal{E}_{i}\leq 2\%$ in order
to provide the conditions for a high irradiation dose being delivered to the
tumor, while sparing neighboring tissues. In Ref. \cite{BKh} it was shown
that such a required beam of laser accelerated ions can be obtained using a
double layer target. Extensive computer simulations of this target were
performed in Ref. \cite{Es-2002} and the results of experimental studies of
this ion acceleration mechanism are presented in Ref. \cite{Nat}.

\subsubsection{Ion acceleration during plasma expansion into vacuum}

Ion acceleration during the collisionless plasma expansion into vacuum
appears to be one of the most obvious mechanisms of the ion acceleration 
\cite{GP}. In particular, it has been considered as one of the possible
acceleration mechanisms in space plasmas \cite{ACR}. When the electrons that
have been heated and tend to expand overtake the ions in a relatively small
volume, the electric neutrality of the plasma breaks and the generated
electric field induces the ion motion. Although a velocity of the bulk ion
and electron motion is of the order of the ion acoustic speed, $v_{s}=\sqrt{%
T_{e}/m_{p}}$, a small fraction at the plasma front gains energy
efficiently. Here $T_{e}$ is the electron temperature and $m_{p}$ is the ion
(proton) mass. Under the most favorable conditions the ions achieve a
kinetic energy which corresponds to an ion velocity of the order of the
electron thermal velocity, i.e. the maximum ion energy can be of the order
of $m_{p}\mathcal{E}_{e}/m_{e}$. We notice here that for the electron
distributions with the energy cut-off this conclusion requires careful
analysis (see Refs. \cite{M-T,BEKTF})

In the limit when the electron energy is relativistic, in order to analyze
the ion motion one should use the equations of relativistic hydrodynamics, $%
\nu _{\alpha} T_{\mu} ^{\nu} = 0$ with $T_{\mu}^{\nu} $ being the
energy-momentum tensor, 
\begin{equation}
\partial _{\mu }\left( {nu^{\mu }}\right) =0,
\end{equation}
\begin{equation}
\mathcal{W}u^{\mu }\partial _{\mu }u^{\nu }=-\left( {\delta ^{\mu \nu
}-u^{\mu }u^{\nu }}\right) \partial _{\mu }P.
\end{equation}
Here $u^{\mu} $ is the four-dimensional velocity vector, $P$ is the
pressure, n is the density in the proper frame of reference, $\mathcal{W} =
P + \varepsilon $ is the enthalpy with $\varepsilon $ being the internal
energy density.

The self-similar plasma motion depending on the variable $\chi =x/t$ is
described by a system of ordinary differential equations 
\begin{equation}
\frac{u\chi -1}{(u-\chi )(1-u^{2})}u^{\prime }-(\ln n)^{\prime }=0,
\end{equation}
\begin{equation}
\mathcal{W}\frac{u-\chi }{1-u^{2}}u^{\prime }-(u\chi -1)P^{\prime }=0,
\end{equation}
with a prime denoting a differentiation with respect to $\chi $ and $u=v/c$.
Here we use the relativistic equation of state of an ideal gas, 
\begin{equation}
\mathcal{W}=c^{2}\frac{K_{3}\left( {m_{e}c^{2}/T_{e}}\right) }{K_{2}\left( {%
m_{e}c^{2}/T_{e}}\right) },
\end{equation}
\begin{equation}
P=nT_{e},
\end{equation}
where $K_{n}(x)$ are the modified Bessel functions. In the case of $T_{e}=$%
constant we find 
\begin{equation}
u=\frac{c_{s}+c\eta }{c+c_{s}},
\end{equation}
\begin{equation}
n=n_{0}\left( \frac{c_{s}-c\eta }{c+c_{s}}\right) ^{c/c_{s}}
\end{equation}
with $c_{s}$ being the relativistic speed of sound, 
\begin{equation}
c_{s}=\sqrt{\frac{T_{e}}{m_{e}}\frac{K_{3}(m_{e}c^{2}/T_{e})}{%
K_{2}(m_{e}c^{2}/T_{e})}}.
\end{equation}
In the ultra-relativistic limit the energy spectrum of fast ions has a
power-law form, 
\begin{equation}
\frac{d\mathcal{N}_{p}(\mathcal{E}_{i})}{d\mathcal{E}_{i}}\propto \mathcal{E}%
_{i}^{-2c^{2}/c_{s}^{2}}.
\end{equation}

\subsubsection{Radiation pressure dominated regime of the ion acceleration}

\begin{figure}[tbp]
\includegraphics[width=9cm]{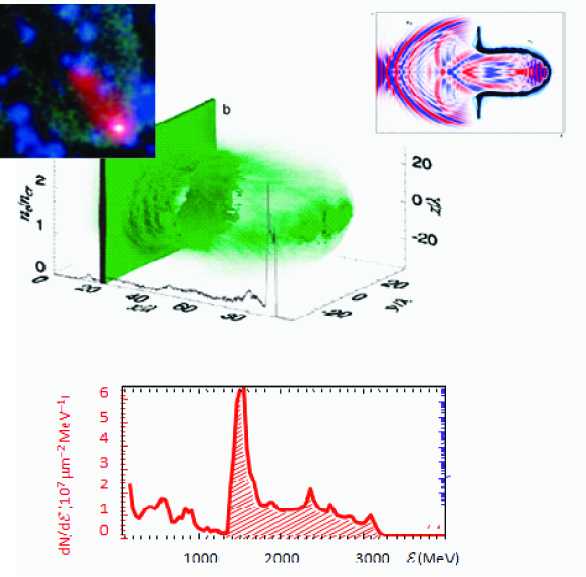} 
\caption{ Results of 3D PIC simulations of the PPDA ion acceleration regime.
a) The electromagnetic pulse forms a cocoon confining the EMW energy. The
right inset shows a cocoon seen in the plasma density and an EMW
distribution obtained with the 2D PIC simulation. In the left inset we see a
cocoon formed by the Black Widow pulsar (Pic. NASA). b) Quasi-monoenergetic
ion spectrum.}
\label{fig:07}
\end{figure}

A regime of ion acceleration that exhibits very favorable properties has
been identified in Ref. \cite{RPDA}. Among the wide variety of ion
acceleration mechanisms realized in the laser-plasma interaction, the
radiation pressure dominated ion acceleration (RPDA) has the highest
efficiency. In the RPDA ion accelerator the laser pulse radiation pressure
pushes forward the irradiated region of a thin foil as a whole. In the
relativistic limit, when the electrons and ions move together with the same
velocity due to a smallness of the electron to ion mass ratio, the ion
kinetic energy is by a factor $m_{i}/m_{e}$ times higher than the electron
energy. In this case the laser pulse interacts with an accelerated foil like
with a relativistic co-propagating mirror. The electromagnetic radiation
reflected back by the relativistic mirror has almost negligible energy
compared to the energy in the incident laser pulse, i.e. the laser energy is
almost completely transformed into the energy of fast ions. In Fig. \ref%
{fig:07} we show results of 3D PIC simulations of this ion acceleration
regime. In the course of the interaction with a thin overdense plasma slab
the multi-petawatt laser pulse forms a cocoon confining the EMW energy, thus
increasing the coupling of the electromagnetic wave with the target (see
frame a) and the 2D inset). The ions accelerated beyond the GeV energy level
have a quasi-monoenergetic spectrum (Fig. \ref{fig:07}b). We notice that a
combination of the RPDA mechanism with the use of double layer targets can
substantially increase the ion acceleration efficiency as demonstrated in
Ref. \cite{lif}. 

The equations of the irradiated foil motion can be cast into the form \cite%
{PB-07}: 
\begin{equation}
\frac{dp_{i}}{dt}=\mathcal{P}d\sigma _{i}\mathbf{\;},  \label{eq1}
\end{equation}
where $p_{i}$ is a momentum of the foil element, $d\sigma _{i}$ is a vector
normal to the foil, the index $i=1,2,3$, and $\mathcal{P}$ is the
relativistically invariant pressure. In the frame of reference co-moving
with the foil the radiation pressure is equal to $\mathcal{P}=E_{M}^{2}/2\pi 
$, with $E_{M}$ being the EMW amplitude. In the laboratory frame of
reference we have $E_{0}^{2}=E_{M}^{2}\left( \omega _{0}/\omega _{M}\right)
^{2}$, where $\omega _{0}$ and $\omega _{M}$ are the wave frequency in the
laboratory and boosted frames. They are related to each other as $\omega
_{M}/\omega _{0}=\sqrt{\left( 1-\beta \right) /\left( 1+\beta \right) }$.
Introducing the Lagrange variables $\eta $ and $\xi $, related to the Euler
coordinates as $x=x(\eta ,\xi ,t)$, $y=y(\eta ,\xi ,t)$, $z=z(\eta,\xi,t)$,
we find that the vector normal to the foil surface element is given by $%
d\sigma _{i}=\varepsilon _{ijk}dx_{j}dx_{k}$. Here $dx_{j}$ are the vectors
directed along the $i$-axes, $\varepsilon _{ijk}$ is the fully antisymmetric
unity tensor, and a summation over repeated indices is assumed. Using these
relationships we can find the equations of foil motion 
\begin{equation}
\frac{\partial p_{i}}{\partial t}=\frac{\mathcal{P}}{\nu _{0}}\varepsilon
_{ijk}\frac{\partial x_{j}}{\partial \eta }\frac{\partial x_{k}}{\partial
\xi },
\end{equation}
\begin{equation}
\frac{\partial x_{i}}{\partial t}=c\frac{p_{i}}{\sqrt{%
m_{p}^{2}c^{2}+p_{k}p_{k}}}.
\end{equation}
Here $\nu _{0}=n_{0}l_{0}$ is the initial surface density, $%
p_{i}=(p_{x},p_{y},p_{z})$ is the momentum, $x_{i}=(x,y,z)$ is the foil
element coordinate and, the index, $m_{p}$ is the ion mass. In the
nonrelativistic limit for constant pressure $\mathcal{P}$, this system is
reduced to the equations obtained in Ref. \cite{Ott}.

When a planar foil is irradiated by a normally incident EM pulse, the ions
achieve the energy 
\begin{equation}
\gamma _{i}=1+\frac{2w^{2}}{1+2w},
\end{equation}
where $w$ is the normalized fluence, 
\begin{equation}
w=\int_{-\infty }^{t-x/c}\frac{E_{0}^{2}(\psi )}{2\pi n_{0}lm_{i}c}d\psi .
\end{equation}
In the limit $w\gg 1$ the resulting ion energy is equal to the ratio of the
laser pulse energy, $\mathcal{E}_{las}$ , to the total number of accelerated
ions, $N_{tot}$, i.e. $\gamma _{i}\approx \mathcal{E}%
_{las}/m_{i}c^{2}N_{tot} $. As an example, we consider a solid density foil, 
$n_{0}=10^{24}$cm$^{-3}$, of 1$\ \mu $m thickness irradiated by a laser
pulse with a transverse size of 100$\ \mu $m. For the laser pulse energy of
the order of 200\ kJ we find that the accelerated ion energy is equal to 1$\ 
$TeV with a total ion number of 10$^{12}$. The ion acceleration length in
this case is approximately equal to $l_{acc}\approx 0.5~$km.

In order to achieve high values of the ion bunch luminosity it is highly
desirable to decrease the transverse bunch size. This can be achieved by
modulating the density inside the foil, e.g. by a properly modulated laser
pulse. The analysis of the linearized equations of the foil motion
demonstrates the exponential growth of the modulations 
\begin{equation}
x_{i}^{(1)}(\alpha ,\beta ,t)\propto \exp \left[ {\left( {\frac{t}{\tau _{RT}%
}}\right) ^{1/3}-iq\eta -ir}\xi \right] ,  \label{eq34}
\end{equation}
where 
\begin{equation}
\tau _{RT}=\omega _{0}^{-1}\frac{(2\pi )^{3/2}R_{0}^{1/2}}{6\left( {%
q^{2}+r^{2}}\right) ^{3/2}\lambda _{0}^{2}}.  \label{eq37}
\end{equation}
and \cite{PB-07} 
\begin{equation*}
R_{0}=\frac{E_{0}^{2}}{2\pi n_{0}l\omega _{0}^{2}}.
\end{equation*}
This opens a way for focusing the accelerated ions onto a narrow spot with
lower limit given by the foil thickness. Using these results we can estimate
the RPDA accelerated ion bunch luminosity as 
\begin{equation}
\mathcal{L}=10^{35}\left( \frac{f}{10KHz}\right) \left( \frac{N_{tot}}{%
10^{12}}\right) ^{2}\left( \frac{10^{-4}cm}{\sigma _{\bot }}\right) ^{2}\ 
\frac{1}{\mathrm{cm}^{2}\mathrm{s}}.
\end{equation}

A first indication of the RPDA - regime has been obtained in the experiments 
\cite{Kar}, when a thin foil target has been irradiated by a laser pulse
with an intensity approaching 10$^{20}$W/cm$^{2}$.

\section{Mini-black-holes on Earth}

\label{MBHE}

We may see that when LWFA and RPDA accelerators will reach 100 GeV and TeV
particle energies, which corresponds to the energy range of interest for
high energy physics, laser accelerators may be considered as a source of
ultrarelativistic particle beams with parameters comparable to those, which
are produced by standard accelerators.

As an example for the problems in the field of high energy physics and
astrophysics which may be explored with laser accelerators of charged
particles, we note the mini-black-hole detection. In the general relativity
theory black holes play a fundamental role. The Einstein equation \cite%
{LLTF,WMTW}, 
\begin{equation}
R_{\mu \nu }-\frac{1}{2}g_{\mu \nu }R=-\frac{8\pi }{m_{p}^{2}}T_{\mu \nu },
\end{equation}
where $m_{p}^{2}=1/G$ is the square of the Planck mass, and $T_{\mu \nu }$
is the energy-momentum tensor (the units $\hbar =c=1$ are used), has a
Schwarzschild solution for the interval: 
\begin{equation}
ds^{2}=g_{\mu \nu }dx^{\mu }dx^{\nu }=-\eta (r)dt^{2}+dr^{2}/\eta
(r)+r^{2}d\Omega ^{2}
\end{equation}
with 
\begin{equation}
\eta (r)=1-(2/m_{p}^{2})M/r.
\end{equation}
Here $M$ and $\Omega $ are the object mass and the surface element in the 3D
space. The metric given by this interval has a singularity at $r$ equal to
the Schwarzschild radius, $R_{BH}=2M/m_{p}^{2}$. For an object with a mass
of the order of the solar mass the black hole radius is equal to 2\ km. As
it was noted above, a black hole with the size of about the Planckian
length, 10$^{-33}$cm, has the mass $m_{p}=10^{-5}$g, which corresponds to an
energy approximately equal to 10$^{19}$GeV.

The situation may change, if our world's dimension is higher than 3. In
accordance with modern quantum field theory \cite{ArH}, our world may have
higher dimensions $(d+3)$. The additional dimensions are compactified in a
sufficiently small scale, $R_{comp}$. Gravitational interaction is present
in the whole space due to its universal character. At the small scale for $%
r\ll R_{comp}$, the gravitational potential of the field produced by an
object with mass $M$ behaves as $\phi (r)=M_{f}^{d+2}M/r^{(1+d)}$. A
constant $M_{f} $ characterizes the gravitational interaction in the small
scale limit. In the limit of large scale compared with $R_{comp}$, i.e. for $%
r\gg R_{comp}$ we have the expression $\phi
(r)=M/m_{p}^{2}r=M/rR_{comp}^{d}M_{f}^{d+2}$. It yields a relationship
between $m_{p}$, $R_{comp}$, and $M_{f}$ being $%
m_{p}^{2}=M_{f}^{d+2}R_{comp} $. The solution of the Einstein equation in
space with d+3 dimension gives for the interval the formula 
\begin{equation}
ds^{2}=-\eta (r)dt^{2}+dr^{2}/\eta (r)+r^{2}d\Omega _{d+3}^{2},
\end{equation}
where $\Omega _{d+3}$\ is the surface element and the metric element is
equal to 
\begin{equation}
\eta (r)=1-(R_{BH}/r)^{d+1}.
\end{equation}
This results in the expression for the black hole radius 
\begin{equation}
R_{BH}^{d+1}=(2/(d+1))M_{f}^{-(d+1)}M/M_{f}.
\end{equation}

For the constant $M_{f}$ of the order of 1 TeV the black hole radius is
equal to $R_{BH}\approx 10^{-4}\ $fm (here 1\ fm=10$^{-13\ }$cm). A
probability of a black hole creation is proportional to the cross section of
this process \cite{MBH} 
\begin{equation}
\sigma _{BH}=\pi R_{BH}^{2}.
\end{equation}
In the collision of 7 TeV proton bunches with a luminosity of the order of
the LHC luminosity, $\mathcal{L}=10^{34}$cm$^{-2}$s$^{-1}$, it is expected
that approximately 10$^{9}$ mini-black-holes may be detected per year. The
created mini-black-holes are to be detected by their emission of
electromagnetic radiation and of elementary particles according to the
Hawking mechanism. At the end of its evolution the black hole is thought be
strings.

The TeV range laser accelerator of charged particles can generate 10$^{6}$
mini-black-hole per year, when its repetition rate is 1 Hz.

\section{Flying Mirror Concept of the Electromagnetic Wave Intensification}

\label{FMC}

An electromagnetic wave reflected off a moving mirror undergoes frequency
multiplication and corresponding increase in the electric field magnitude.
The multiplication factor $(1+\beta _{M})/(1-\beta _{M})$ is approximately
proportional to the square of the Lorentz factor of the mirror, $\gamma
_{M}=1/\sqrt{1-\beta _{M}^{2}}$, making this effect an attractive basis for
a source of powerful high-frequency radiation. Several ways have been
suggested to extremely high intensity (see articles, Refs. \cite%
{BET-03,ARUT,NAU}, \cite{MTB-06,PPR} and literature quoted in). A specular
reflection by a sufficiently dense relativistic electron cloud as suggested
in Refs. \cite{LAND}. The reflection at the moving ionization fronts was
studied in Refs. \cite{IONS}.

\begin{figure}[tbp]
\includegraphics[width=9cm]{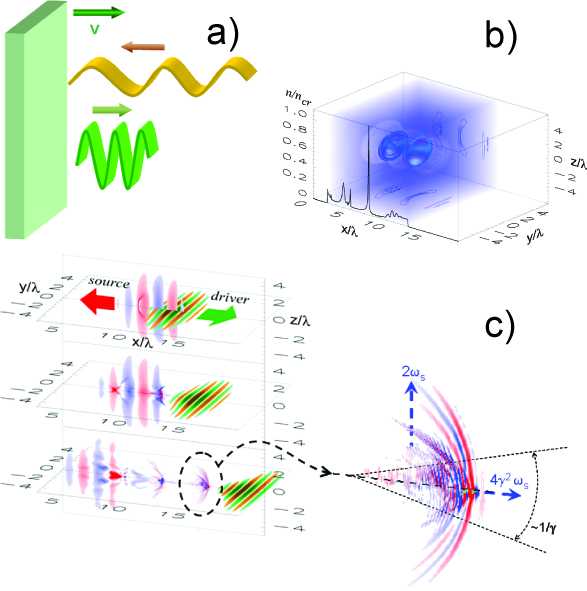} 
\caption{Flying Mirror Concept. a) The reflection of EMW at the relativistic
mirror results in a frequency upshifting and compression of the wave. b)
Paraboloidal modulations of the electron density in the plasma wake wave. c)
The electric field pattern of the laser pulse driver and of the reflected
EMW. Inset: The reflected electromagnetic pulse frequency is upshifted, it
is focused and its intensity increases.}
\label{fig:08}
\end{figure}

Here we consider the \textquotedblleft flying mirror\textquotedblright \
concept\cite{BET-03}. It uses a fact that at optimal conditions, the dense
shells formed in the electron density in a strongly nonlinear plasma wake,
generated by a short laser pulse, reflect a portion of a counter-propagating
laser pulse. In the wake wave generated by the ultrashort laser pulse
electron density modulations take the form of a paraballoid moving with the
phase velocity close to the speed of light in vacuum \cite{PAR}. At the wave
breaking threshold the electron density in the nonlinear wake wave tends
towards infinity. The formation of peaked electron density maxima breaks the
geometric optics approximation and provides conditions for the reflection of
a substantially high number of photons of the counterpropagating laser
pulse. As a result of the electromagnetic wave reflection from such a
"relativistic flying mirror", the reflected pulse is compressed in the
longitudinal direction, which is a consequence of frequency upshifting. The
paraboloidal form of the mirrors leads to a reflected wave focusing into the
spot with the size determined by the shortened wavelength of the reflected
radiation (see Fig. \ref{fig:08}). This mechanism allows to generate
extremely short, femto-, atto-, zepto-second duration pulses of coherent
electromagnetic radiation with extremely high intensity, which pave the way
for studying such nonlinear quantum electrodynamics effects as the
electron-positron pair creation and nonlinear refraction in vacuum. 

The key parameter in the problem of Flying Relativistic Mirror (FRM) is the
wake wave gamma factor, $\gamma _{ph,W}$. According to the special theory of
relativity \cite{STR}, the frequency of the electromagnetic wave reflected
from FRM increases by a factor approximately equal to $4\gamma _{ph,W}^{2}$.
A number of back reflected photons is proportional to $\gamma _{ph,W}^{-3}$
(for details see Ref. \cite{PAN}), which results in the reflected light
intensification \cite{BET-03} 
\begin{equation}
I_{r}/I_{0}\approx \gamma _{ph,W}^{3}(S/\lambda _{0})^{2},  \label{eq:intFM}
\end{equation}
where $S$ is the transverse size of the laser pulse incident on the FRM. The
reflected pulse power increases as $\mathcal{P}_{r}=\mathcal{P}_{0}\gamma
_{ph,W}$.

Using the expression for the reflected pulse intensity (\ref{eq:intFM}), we
obtain that the interaction of two laser pulses with energies 10 kJ and 30
J, respectively, counterpropagating in a plasma with a density $\approx
10^{18}$cm$^{-3}$ can result in a light intensification of up to $\approx
10^{28}$W\ cm$^{-2}$. This corresponds to the generation of an electric
field with a value close to the nonlinear quantum electrodynamics (QED)
limit, $E_{QED}=m_{e}^{2}c^{3}/e\hbar $, when electron-positron pairs can be
created in vacuum. This QED electric field is also called the "Schwinger
field".

Experiments utilizing the electromagnetic pulse intensified with the FRM
technique may allow studying regimes of super-Schwinger fields, when $%
E>E_{QED}$. This may be possible because the light reflected by the
parabaloidal FRM is focused into a focus spot moving with a relativistic
velocity and is well collimated within an angle $\approx 1/\gamma _{ph,W}$ 
\cite{BET-03}. The wave localization within the narrow angle corresponds to
the fact that the wave properties are close to the plane wave properties to
the extent of the smallness of the parameter $1/\gamma _{ph,W}$. In this
case the second Poincare invariant of the electromagnetic field, $%
B^{2}-E^{2} $, has a value of the order of $E^{2}/\gamma _{ph,W}^{2}$.
Therefore the electric field amplitude in the reflected electromagnetic wave
can exceed the Schwinger limit by $\gamma _{ph,W}$ times. We note that a
tightly focused electromagnetic wave cannot have an amplitude above $E_{QED}$%
, due to the electron-positron pair creation \cite{NAR} when $E\rightarrow
E_{QED}$ leads to the depletion of the electromagnetic wave \cite{BFP}.

As it was shown above, the critical power for mutual focusing of two
counterpropagating EMW is equal to $\mathcal{P}_{cr}=2.5\times 10^{24}$W,
which is beyond the reach of existing and planned lasers. Fortunately, if we
take into account that the radiation reflected by the FRM has a shortened
wavelength $\lambda _{r}=\lambda _{0}/4\gamma _{ph,W}^{2}$ and that its
power is increased by a factor $\gamma _{ph,W}$, we may find that for $%
\gamma _{ph,W}=30$, i.e. for a plasma density $\approx 3\times 10^{17}$cm$%
^{-3}$, nonlinear vacuum properties can be seen for laser light the incident
on the FRM with a power of about 10 PW. This makes the FRM concept
attractive for the purpose of studying nonlinear quantum electrodynamics
effects.

Within the framework of the Flying Mirror concept, it has been demonstrated 
\cite{BET-03} that the wavelength of the laser pulse, which has been
reflected and focused at the wake plasma wave, becomes shorter by a factor $%
4\gamma _{ph}^{2}$ and its power increases by a factor $2\gamma _{ph}$. From
this it follows that nonlinear QED vacuum polarization effects are expected
to be observable for 50 PW 1-$\mu $m lasers.

A demonstration of the Flying Mirror concept has been accomplished in the
experiments of Ref. \cite{KP}. Two beams of terawatt laser radiation
interacted with an underdense plasma slab. The first laser pulse excited the
nonlinear wake wave in a plasma with parameters required for the wave
breaking, which has been seen in the quasi-mono-energetic electron
generation and in the stimulated Raman scattering. The second
counter-crossing laser pulse has been partially reflected from the
relativistic mirrors formed by the wake plasma wave. We detected the
electromagnetic pulses with a duration of femtoseconds and wavelengths from
7 nm to 15 nm. These results demonstrate the feasibility of constructing
sources of coherent X-ray radiation with the parameters that are tunable in
a broad range.

\section{Reconnection of Magnetic Field Lines \& Vortex Patterns}

The term \textquotedblright magnetic field line
reconnection\textquotedblright\ refers to a broad range of problems that are
of interest for space and laboratory plasmas. The results of theoretical and
experimental studies of magnetic reconnection have been reviewed in many
papers and monographs~\cite{ACR,SYROV,BISKAMP}. As it concerns relativistic
laser plasmas, earlier a conclusion was made in Ref. \cite{ASK} about the
important role of the generation of magnetic fields by fast electron
currents and their reconnection in the relativistic laser-matter interaction
regime. The experiments conducted in Refs. \cite{Nil} revealed magnetic
reconection phenomena in laser plasmas, when two high power laser beams
irradiated a thin foil target.

Processes of reconnection are accompanied by an ultra fast magnetic energy
release, which is transformed into different forms, such as internal plasma
energy, radiation and fast particles.

\subsection{Dimensionless parameters describing the relative roles of
nonlinear, dissipative and Hall effects}

\label{S.DIM}

Inside current sheets, which are basic entities in the reconnection process,
as well as in the vicinity of shock wave fronts, the effects of dissipation
and of nonlinearity play a crucial role being comparable in magnitude.
Together with the Hall effect, which leads to the appearance of small scale
structures, these effects violate the freezing of magnetic field in plasma
motion.

Magnetic field is frozen in a plasma in the limit of a large Lundquist
number $S\rightarrow \infty $. It obeys the equation: 
\begin{equation}
\partial _{t}\mathbf{B}=\mathbf{v}\times \mathbf{B},
\end{equation}
which corresponds to the conservation of magnetic field flux through a
contour moving with the plasma. The dimensionless parameter $S$ is equal to
the ratio of two characteristic time scales: the magnetic diffusion time $%
\tau _{\sigma }=l^{2}/\nu _{m}$ and a typical time $\tau _{A}=l/v_{A}$ that
it takes for an Alfv\`{e}n wave to propagate along the distance $l$; $S=\tau
_{\sigma }/\tau _{A}$. Here the magnetic diffusivity is $\nu _{m}=c^{2}/4\pi
\sigma $, where $\sigma $ is the electric conductivity of the plasma, and $%
v_{A}=|\mathbf{B}|/\sqrt{4\pi \rho }$ is the Alfv\`{e}n wave velocity.

In the vicinity of the zero point the scale of the field nonuniformity $l$
equals to the distance $r$ from the zero point. The magnetic field and hence
the Alfv\`{e}n velocity are proportional to $r$: $|\mathbf{B}|=hr$, $%
v_{A}=hr/\sqrt{4\pi \rho }\equiv \Omega _{A}r$. Here $h$ is a typical value
of the gradient of the magnetic field.

The measure of the significance of nonlinear effects is given by the ratio $%
\delta B/hl$ between the magnitude of the magnetic field perturbation $%
\delta B$ and the background magnetic field $B=hl$. This ratio depends on
the distance from the null point, due to both the nonuniformity of the
background magnetic field $B$ and to the change of the MHD wave amplitude in
the course of its propagation.

If the plasma is pinched by a the quasi-cylindrical electric current $I$
with a radius of the order of $r$, the value of the magnetic field at its
boundary is approximately $\delta B=2I/cr$. The dimensionless ratio $\delta
B/hl$ is equal to one for $r\approx r_{m}=\sqrt{I/hc}$. If the electric
current has the form of a quasi-one-dimensional slab pinch, and if the
pinching occurs in the direction of its small size, the characteristic value
of the magnetic field perturbation is constant: $\delta B=B_{\Vert }$ and
the ratio $\varepsilon $ becomes of order unity at the distance $%
r_{A}=B_{\Vert }/h$. In the approximation of small amplitude perturbations,
these two types of pinching correspond to the effects of the propagation of
magnetoacoustic and of Alfv\`{e}n waves, respectively. The magnetoacoustic
waves focus towards the null line, while the energy of the Alfv\`{e}n waves
accumulate near the magnetic field separatrices. The values $r_{m}$ and $%
r_{A}$ determine the size of the region, where the magnetoacoustic wave and,
respectively, the Alfv\`{e}n one become nonlinear.

The dimensionless parameters 
\begin{equation}
(r_{m}/r_{\sigma })^{2}=I\Omega _{A}/c\nu _{m}\equiv L_{m},  \label{eq:4.3}
\end{equation}
\begin{equation}
(r_{A}/r_{\sigma })^{2}=B_{\Vert }^{2}\Omega _{A}/h^{2}\nu _{m}\equiv L_{A}
\label{eq:4.4}
\end{equation}
determine the relative role of the dissipation and of the nonlinearity
effect in the course of the current sheet formation due to finite amplitude
perturbations of the magnetoacoustic and of the Alfv\`{e}n wave type,
respectively.

Now we discuss the relationship between the dimensionless parameter $L_{m}$
and the current sheet parameters obtained in the framework of the Sweet --
Parker model \cite{SWEET,PARKER}. In this model it is supposed the current
sheet has a width $b$ and a thickness $a$ with $b\gg a$. The plasma flows
into the current sheet with a velocity $v_{\mathrm{in}}\approx \nu _{m}/a$
and exits through its narrow edges with a velocity $v_{\mathrm{out}}$, which
is of the order of the Alfv\`{e}n wave velocity, $v_{A}\approx \Omega _{A}b$%
. From mass conservation we obtain $v_{\mathrm{in}}b=av_{\mathrm{out}}$.
From this it follows that the thickness of the current sheet is equal to 
\begin{equation}
a=\sqrt{\nu _{m}/\Omega _{A}},
\end{equation}
i.e. of order $r_{\sigma }$. Estimating the current sheet width as $r_{m}$,
we find that the ratio of its width to its thickness is 
\begin{equation}
b/a=\sqrt{I\Omega _{A}/hc\nu _{m}}\equiv \sqrt{L_{m}}.
\end{equation}
Thus, the condition for the formation of a wide current sheet with $b\gg a$
is equivalent to the requirement $L_{m}\gg 1$. Similarly, current sheets are
formed in the vicinities of the magnetic field separatrices when $L_{A}\gg 1$%
.

Considering the case when the Hall effect, i.e. the electron inertia, plays
a dominant role in the reconnection process, we define the dimensionless
parameter which measures the role of the Hall effect as $\tilde{\alpha}%
=\alpha h/\Omega _{A}l\equiv c/\omega _{pi}l$. When the length $r_{H}$, at
which the Hall effect starts to be important, is larger than the current
sheet thickness, $r_{H}/a=\alpha h/\sqrt{\nu _{m}\Omega _{A}}>1$, the
effects of dispersion lead to the formation of small scale structures. In
the limit $r_{H}/b=\alpha h\sqrt{ch/I}/\Omega _{A}\gg 1$ the pattern of the
plasma flow is completely determined by the Hall effect. Similar to the way
used to define the parameters $L_{m}$ and $L_{A}$, we define the
dimensionless parameter 
\begin{equation}
L_{H}=(b/r_{H})^{2}=E^{2}c^{2}\Omega _{a}/h^{4}\alpha ^{2}\nu _{m}\equiv
E^{2}\omega _{pi}^{2}/h^{2}\nu _{m}\Omega _{a}.
\end{equation}
When $L_{H}\gg 1$, nonlinear effects are much stronger than the Hall effect.

\subsection{Current Sheet}

\begin{figure}[tbp]
\resizebox{0.5\textwidth}{!}{\includegraphics{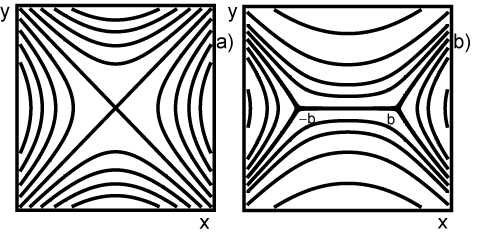}}
\caption{Magnetic field pattern in the vicinity of the X-line (a). Current
layer formed in the vicinity of the X-line (b).}
\label{fig:09}
\end{figure}

In a simple 2D configuration the current sheet is formed in the magnetic
field described by a complex function $B(x,y)=B_{x}-iB_{y}=h\zeta $ of a
complex variable $\zeta =x+iy$. The magnetic field vanishes at the
coordinate origin. The magnetic field lines lie on the surfaces of constant
vector potential, $A(x,y)=\mathrm{Re}\{h\zeta ^{2}/2\}$. They are hyperbolas
as we can see in Fig. \ref{fig:09} a). This is a typical behaviour of the
magnetic field lines in the vicinity of null lines (they are the so called
X-lines) in magnetic configurations. \ 
Under finite time perturbations the magnetic X-line evolves to the magnetic
configurations of the form $B=h(\zeta -b)^{1/2}$, which desribes the
magnetic field created by thin current sheet between two points $\pm b$ \cite%
{SYROV}. The magnetic field lines lie on the constant surfaces of 
\begin{equation}
A(x,y)=\frac{h}{2}\mathrm{Re}\left\{ \zeta \sqrt{\zeta ^{2}-b^{2}}-\mathrm{%
Log}\left[ \zeta +\sqrt{\zeta ^{2}-b^{2}}\right] \right\} .
\end{equation}
They are shown in Fig. \ref{fig:09} b. The width of the current layer $b$ is
determined by the total electric current $I$ inside, and by the magnetic
field gradient, $h$. It is equal to

\begin{equation}
b=\sqrt{4I/hc.}
\end{equation}

\begin{figure}[tbp]
\resizebox{0.45\textwidth}{!}{\includegraphics{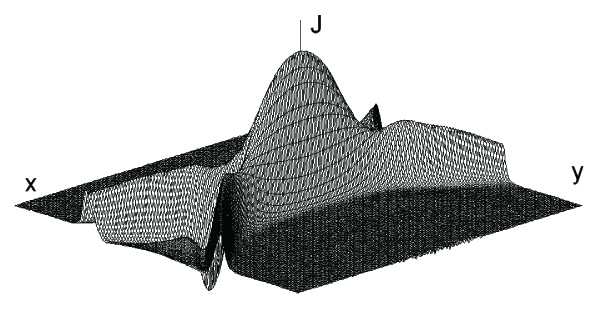}}
\caption{Electric current density distribution inside the current sheet 
\protect\cite{RecPLA}.}
\label{fig:10}
\end{figure}

In the strongly nonlinear stage of the magnetic field and plasma evolution a
quite complex pattern in the MHD flow in the nonadiabatic region near the
critical point can be formed, with shock waves and current sheets. In Fig. %
\ref{fig:10} we show the results of the dissipative magnetohydrodynamics
simulations of the current sheet formation near the X-line. \ 

\subsection{Magnetic Reconnection in Collisionless Plasmas}

When the Hall effect is dominant, i.e. the electron inertia determines the
relationship between the electric field and the electric current density
carried by the electron component, the magnetic field evolution is described
by the equation (see \cite{EMHD,BPS-92}) 
\begin{equation}
\partial _{t}(\mathbf{B}-\Delta \mathbf{B})=\nabla \times \left[ \left(
\nabla \times \mathbf{B}\right) \times (\mathbf{B}-\Delta \mathbf{B})\right]
,  \label{eq:EMHD}
\end{equation}
which corresponds to the condition of generalized vorticity, $\mathbf{\Omega 
}=\mathbf{B}-\Delta \mathbf{B}$, be frozen into the electron component
motion with the velocity $\mathbf{v}_{e}=c\nabla \times \mathbf{B/}4\pi
n_{0}e$. Here the space scale is chosen to be equal to the collisionless
electron skin-depth, $d_{e}=c/\omega _{pe}$, and the time unit is $\omega
_{Be}^{-1}=m_{e}c/eB$. The range of frequencies described by the EMHD
equations is given by $\omega _{Bi}<\omega <\omega _{Be}$.

In the linear approximation Eq.~(\ref{eq:EMHD}) describes the propagation of
whistler waves, for which the relationship between the wave frequency and
the wave vector, is $\omega =|\mathbf{k}|(\mathbf{k}\cdot \mathbf{B_{0}}%
)/(1+k^{2})$. From this relationship it follows that in a weakly
inhomogeneous magnetic field the critical points are the points and lines
where $|\mathbf{B_{0}}|=0$ or/and $(\mathbf{k}\cdot \mathbf{B_{0}})=0$.

The electron inertia effects make the reversed magnetic field configuration
unstable against tearing modes~\cite{FKR,LPV}, which result in magnetic
field line reconnection. The slab equilibrium configuration with a magnetic
field given by $\mathbf{B_{0}}=B_{0z}\mathbf{e_{z}}+B_{0x}(y/L)\mathbf{e_{y}}
$, where $B_{0x}(y/L)$ is the function that gives the current sheet magnetic
field, is unstable with respect to perturbations of the form $f(y)\exp
(\gamma t+ikx)$ with $kL<1$. For this configuration one has $(\mathbf{k}%
\cdot \mathbf{B_{0}})=0$ at the surface $y=0$. The growth rate of the
tearing mode instability is \cite{BPS-92,BASOVA,FRUSTRA} $\gamma \approx
(1-kL)^{2}\Delta ^{\prime 2}/kL^{2}$. 
\begin{figure}[tbp]
\resizebox{0.45\textwidth}{!}{\includegraphics{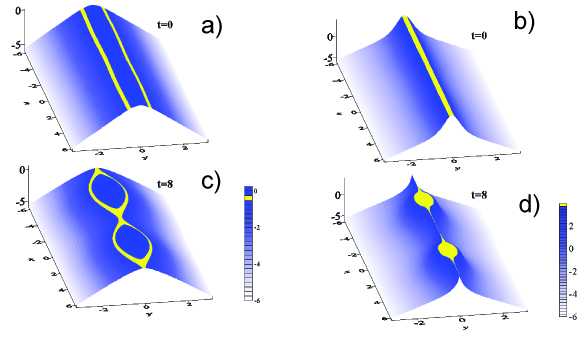}}
\caption{Nonlinear stage of the development of the tearing mode instability
in a current sheet: a) the magnetic field and b) the generalized vorticity
distribution at t=0. The same functions at t=8 in c) and d).}
\label{fig:11}
\end{figure}

In Fig.~\ref{fig:11} the results of a numerical solution of Eq.~(\ref%
{eq:EMHD}) in a 2D geometry with magnetic field $\mathbf{B}(x,y,t)=\left(
\nabla \times a\right) \times \mathbf{e_{\bot }}+b\mathbf{e_{\Vert }}$ are
shown. The unperturbed configuration is chosen to be a current sheet,
infinite in the $x$-direction, that separates two regions with opposite
magnetic field. Both the line pattern of generalized vorticity, $\Omega
=a-\Delta a$, and of the magnetic field show the formation of
quasi--one--dimensional singular distributions in the electric current
density and in the distribution of the generalized vorticity. The magnetic
field topology changes, as is seen from Fig.~\ref{fig:11}.

\subsection{Charged Particle Acceleration}

\begin{figure}[tbp]
\resizebox{0.45\textwidth}{!}{\includegraphics{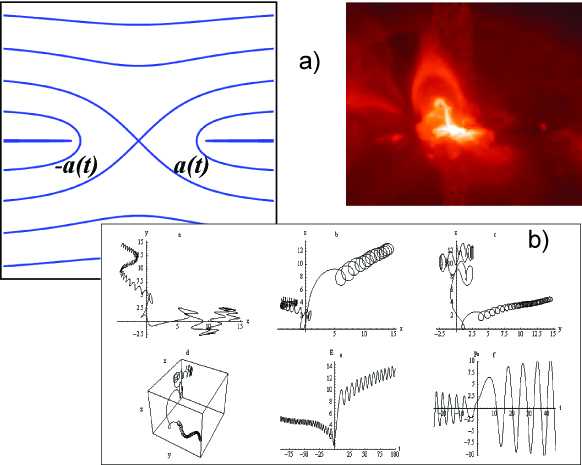}}
\caption{a) The current sheet break up into two parts separated by the
distance $2a(t)$. b) The projections of the trajectory of charged particle
accelerating in the vicinity of the masgnetic X-line. Inset: The solar flare
(YOHKOH image).}
\label{fig:12}
\end{figure}

A fully developed tearing mode results in a current sheet break up into
parts separated by a distance $2a$, as it is illustrated in Fig. \ref{fig:12}%
a (see. Ref. \cite{SYROV} and literature quoted therein). Under the magnetic
field line tension the plasma is thrown out. The model magnetic field
describing this configuration is given by the complex variable function $%
B(\zeta )=B_{0}\varsigma /\sqrt{a^{2}-\varsigma ^{2}}$. The magnetic field
lines lie on the surfaces of constant vector potential, 
\begin{equation}
A(x,y,t)=\mathrm{Re}\left\{ B_{0}\sqrt{a^{2}(t)-\varsigma ^{2}}\right\} .
\end{equation}
Due to a dependence of the function $a$ on time the electric field parallel
to the $z$ axis arrises. It is given by 
\begin{equation}
E(x,y,t)=-\frac{1}{c}\partial _{t}A=-\frac{1}{c}\frac{B_{0}a(t)\dot{a}(t)}{%
\sqrt{a^{2}(t)-\varsigma ^{2}}}.
\end{equation}
In the vicinity of the null line we have a quadrupole structure of the
magnetic $B(\zeta )\approx B_{0}\varsigma /a$ field and a locally
homogeneous electric field, $E\approx \dot{a}B_{0}/c$. 

The magnetic field reconnection, the study of which has been started by
Dungey \cite{Dungey}, on its initial stage had had as a main goal to explain
the generation of suprathermal particles during solar flares and substorms
in the earth's magnetosphere. Despite the simplycity of the formulation of
the problem, it is quite far from a complete solution. Even in the test
particle approximation, which describes the particle motion in the given
magnetic and electric fields, the solution of this problem meets serious
difficulties~\cite{ACR,RCACC}. The reason of that is due to the fact that in
the vicinity of critical points of magnetic configurations the standard
approximations adopted to describe the plasma dynamics are no longer valid.
In such regions the drift approximation, i.e., the assumption that the
adiabatic invariants are constant, can no longer be applied. On the other
hand, the particle spends only a finite time interval in the nonadiabatic
region, since there its motion is unstable. After a finite time interval it
gets out of the nonadiabatic region, and gets into the drift region as it is
seen in Fig. \ref{fig:12}b. Matching the solution described by the particle
trajectories in different regions, we can describe the particle motion and
hence the acceleration near critical points of the magnetic configurations.

Under the conditions of space plasmas, the radiation losses during the
charged particle acceleration in the magnetic reconnection processes are
caused by the backward Compton scattering and by synchrotron losses. A
characteristic time of the synchrotron losses for the electron with energy $%
\mathcal{E}$ is given by the expression 
\begin{equation}
\tau _{B}=\frac{3m_{e}^{4}c^{7}}{2e^{4}B^{2}\mathcal{E}}.
\end{equation}
As it was shown in Ref. \cite{BKORF}, during solar flares this effect limits
the ultrarelativistic electron energy to a value of about several tens of
GeV.

\subsection{Electron Vortices in Collisionless Plasmas}

The vortical fluid motion is well known to be widely present under the
earth's and space conditions. In laser plasmas, when ultra short and high
intensity EMW pulse propagates in the collisionless plasmas, it accelerates
a copious number of relativistic electrons. The electric current of fast
electrons produces quasistatic magnetic field, whose evolution results in
the formation of electron vortex structures. They naturally take a form of
the vortex rows \cite{HMV}, as it is shown in the LHS inset to Fig.\ref%
{fig:13}. A strong magnetic field in the relativistic laser plasma has been
detected experimentaly \cite{BBfield}.

The interacting vortices can be described within the framework of a
two-dimensional theoretical model. By taking $\mathbf{B}$ to be along the $z$%
-axis ($\mathbf{B=}B\mathbf{{e}_{z}}$), and assuming all the quantities to
depend on the $x,y-$coordinates, we obtain from vector equations (\ref%
{eq:EMHD}) one equation 
\begin{equation}
\partial _{t}(\Delta B-B)+\{B,(\Delta B-B)\}=0  \label{eq:8.2D}
\end{equation}
for a scalar function $B(x,y,t)$. Hear 
\begin{equation}
\{f,g\}=\partial _{x}f\ \partial _{y}g-\partial _{x}g\ \partial _{y}f
\end{equation}
are the Poisson brackets. Equation (\ref{eq:8.2D}) is known as the Charney
equation\ \cite{Batch} or the Hasegawa-Mima (HM) \cite{HasMima} equation in
the limit of zero drift velocity. In this case linear perturbations with the
dispersion equation $\omega =|\mathbf{k}|(\mathbf{k}\cdot \mathbf{B_{0}}%
)/(1+k^{2})$ correspond to the Rossby waves, the drift waves or to the
whistler waves, respectively.

Equation (\ref{eq:8.2D}) has a discrete vortex solution, for which the
generalized vorticity is localized at the points $\mathbf{x=x}^{\alpha }$: 
\begin{equation}
\Omega =\Delta B-B=\sum_{\alpha }\kappa _{\alpha }\delta (\mathbf{x-x}%
^{\alpha }(t)).
\end{equation}
Solving this equation we find that the magnetic field is a superposition of
the magnetic fields created at isolated vortices localized at the
coordinates $\mathbf{x}^{\alpha }(t)$: $\,B=\sum_{\alpha }B^{\alpha }$ with $%
~$%
\begin{equation}
B^{\alpha }(\mathbf{x},\mathbf{x}^{\alpha }(t))=-\frac{\kappa _{\alpha }}{{%
2\pi }}K_{0}(|\mathbf{x}-\mathbf{x}^{\alpha }(t)|)\mathbf{.}
\label{eq:8.BVrt}
\end{equation}
Here and below $K_{n}(\xi )$ are modified Bessel functions.

The curves $\mathbf{x}^{\alpha }(t)$ are determined by the characteristics
of equation (\ref{eq:8.2D}). The characteristic equations have the
Hamiltonian form 
\begin{equation}
\kappa _{\alpha }\dot{x}_{i}^{\alpha }=J_{ij}\frac{\partial \mathcal{H}}{%
\partial x_{j}^{\alpha }}=-\frac{1}{2\pi }J_{ij}\sum_{\beta \neq \alpha
}\kappa _{\alpha }\kappa _{\beta }\frac{(x_{i}^{\alpha }-x_{j}^{\beta })}{%
l_{\alpha \beta }^{2}},  \label{eq:8.HpointHM}
\end{equation}
where $J_{ij}$ is the antisymmetric unit matrix. The Hamiltonian is given by 
\begin{equation}
\mathcal{H}=-\sum_{\alpha <\beta }\kappa _{\alpha }\kappa _{\beta
}K_{0}(l_{\alpha \beta })/2\pi .
\end{equation}

In the case of the Euler hydrodynamics, a point vortex is described by $%
(\kappa _{\alpha }/2\pi )\ln \mathbf{|x}-\mathbf{x}^{\alpha }(t)|$, instead
of the expression (\ref{eq:8.BVrt}) which involves the Bessel \ function $%
K_{0}\,(\mathbf{|x}-\mathbf{x}^{\alpha }(t)|\mathbf{)}$. The later results
in the shiealding of the interaction between vortices at large distances. A
typical scale length of the problem under consideration in the case of the
EMHD vortex systems, is equal to the collisionless electron skin-depth, $%
d_{e}=c/\omega _{pe}$.

\begin{figure}[tbp]
\resizebox{0.5\textwidth}{!}{\includegraphics{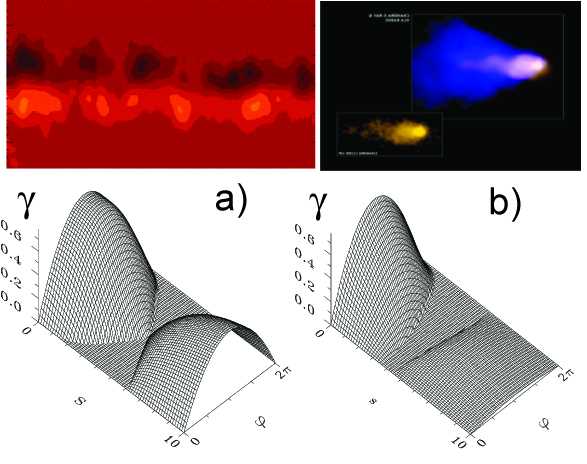}}
\caption{Instability growth rate on $s$ and $q$ for the vortex row described
within the framework of the Euler (a) and Hasegawa-Mima (b) approximations.
Left inset: The vortex row seen in the magnetic field patch distribution in
a plasma behind the laser pulse. Right inset: The Mouse pulsar
(NASA/CXC/SAO, Chandra image of G359.23-0.82 pulsar)\protect\cite{Gaensler}.}
\label{fig:13}
\end{figure}

Considering the problem of the stability of an infinite vortex chain we
assume that all vortices have the same absolute intensity and take. In the
initial equilibrium the vortices have coordinates (Fig. (\ref{fig:13}))

In the case of an antisymmetrical vortex row with $\sigma =1/2$, we expect a
more complicated behavior of the perturbations, compared to that of the
symmetrical configuration. As noted in Lamb's monograph \cite{Lamb}, in
standard hydrodynamics the antisymmetrical von Karman's vortex row is stable
for $q/s\approx 0.281$, where $s$ and $q$ give a distance between the
vortices in the unperturbed vortex row along the $x$ and $y$ coordinates. A
dependence of the instability growth rate on $s$ and $q$ for the vortex row
described within the framework of the Euler hydrodynamics approximation is
shown in Fig. \ref{fig:13}a. 

By direct inspection of the row instability described by the Hasegawa-Mima
equations we can see that for large distance between neighbouring vortices
the antisymmetric vortex row is stable for 
\begin{equation}
3s^{2}/4>q>s/2.
\end{equation}
(see Fig. \ref{fig:13}b).

\subsection{A Role of the Weibel Instability in the Quasistatic and
Turbulent Magnetic Field Generation}

The quasistatic magnetic field generation in relativistic laser plasmas
occurs due to the fast electron beam interaction with the background plasma.
It can be understood in terms of the Weibel instability \cite{Weibel} or in
the generic case, in terms of the electromagnetic filamentation instability.
When the fast electron beam propagates in the plasma, its electric current
is compensated by the current carried by the plasma electrons. A repulsion
of the oppositely directed electric currents results in the electron beam
filamentation and in the generation of a strong magnetic field \cite{Cali}.
An electromagnetic filamentation instability leads to the generation of a
quasistatic magnetic field and is associated with many small-scale current
filaments \cite{Honda}. Each filament consists of a direct and of a return
electric current which repel each other. This produces a strong electric
field, which accelerates the ions in the radial direction. In the long term
evolution, the successive coalescence of the small-scale current filaments
forms a large scale magnetic structure. This process is accompanied by the
reconnection of the magnetic field lines, by the formation of current
sheets, and by strong ion acceleration inside these sheets \cite{Sakai1}.

The filamentation phenomena are of great interest for the explanation of the
quasi-static magnetic field origin in laser plasmas irradiated by
relativistically strong EMW \cite{BAsk}. Counterstreaming electric current
configurations naturally appear in space at the fronts of colliding
electron-positron and electron-ion plasma clouds \cite{Kazimura} as in the
cases of the Galactic Gamma Ray Bursts and in shock waves in supernova
remnants. The filamentation instability generates the magnetic field
required by the theory of the synchrotron afterglow in GRB \cite{Med1}. The
Weibel instability has been invoked as a mechanism of the primordial
magnetic field generation by colliding electron clouds in cosmological
plasmas \cite{Sakai2}.

The filamentation instability developing in the vicinity of shock wave
fronts together with other types of instabilities \cite{TKrll} plays the
role of the source of strong electromagnetic turbulence invoked in the
theoretical models of the Fermi acceleration of cosmic rays \cite%
{Med2,Takabe}. A realization of the Fermi acceleration mechanism of Type A
at the shock wave front is discussed below in Section \ref{Sec:SW}.

\section{Relativistic Rotator}

In Ref. \cite{GG} the antenna mechanism of the pulsar radiation emission has
been proposed. According to this mechanism in the pulsar magnetoshere, which
is a rotating magnetic dipole, the magnetic dipole interaction with a plasma
at the magnetosphere periphery induces strong modulations of the electron
density, an electron density lump. The phase velocity of the electron lump
can be arbitrarily close to the speed of light in vacuum. It is directed
along a circle as illustrated in Fig. \ref{fig:GG}. As a result of the
curvilinear acceleration, the electron lump emits a radiation, whose
properties are similar to the synchrotron radiation \cite{VLG}.

\begin{figure}[tbp]
\resizebox{0.40\textwidth}{!}{\includegraphics{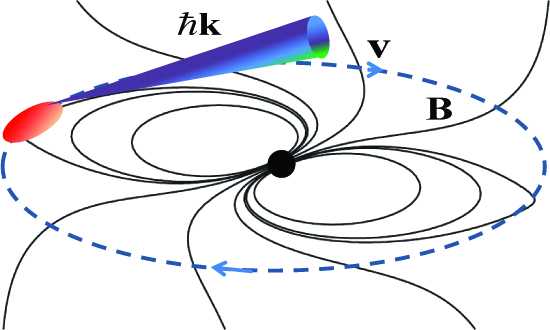} }
\caption{ Schematic pulsar magnetosphere, according to Ref. \protect\cite{GG}%
. A rotating relativistic electron lump emits electromagnetic radiation by
the antenna mechanism. }
\label{fig:14}
\end{figure}

\begin{figure}[tbp]
\resizebox{0.50\textwidth}{!}{\includegraphics{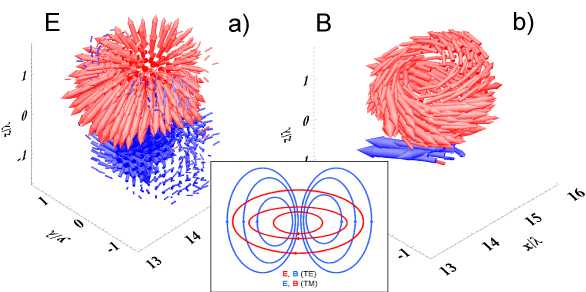} }
\caption{ Structure of electric (a) and magnetic (b) fields inside the EM
relativistic soliton. Inset: The magnetic- and electric-field topology in
the TE (with poloidal magnetic field and toroidal electric field) and in the
TM (with poloidal electric field and toroidal magnetic field) solitons.}
\label{fig:15}
\end{figure}

In the context of Relativistic Laboratory Astrophysics it is remarkable that
the relativistic rotating dipole can naturally be formed in the laser
plasma. Laser-plasma interactions provide an opportunity to reproduce
nonlinear electrodynamics effects under astrophysical conditions in the
laboratory. In Ref. \cite{Afterglow} it is demonstrated that high-power
coherent synchrotron-like radiation can be generated by the relativistic
charge density wave rotating self-consistently inside an
electromagnetic-dipole solitary wave, dwelling in a laser plasma. The
relativistically strong laser pulse can generate relativistic EM subcycle
solitary waves in a plasma \cite{EMSoliton}, as it was indicated by
particle-in-cell (PIC) simulations. An analytical description of solitons of
this type was developed in Refs. \cite{Kozlov}. Figure \ref{fig:15} presents
the structure of electric and magnetic fields inside the soliton \cite%
{3Dsoliton}. The soliton ressembles an oscillating or rotating electric
dipole. The toroidal magnetic field, shown in Fig. \ref{fig:15}, indicates
that, besides the strong electrostatic field, the soliton also has an
electromagnetic field. The electrostatic and electromagnetic components in
the soliton are of the same order of magnitude. 

\begin{figure}[tbp]
\resizebox{0.50\textwidth}{!}{\includegraphics{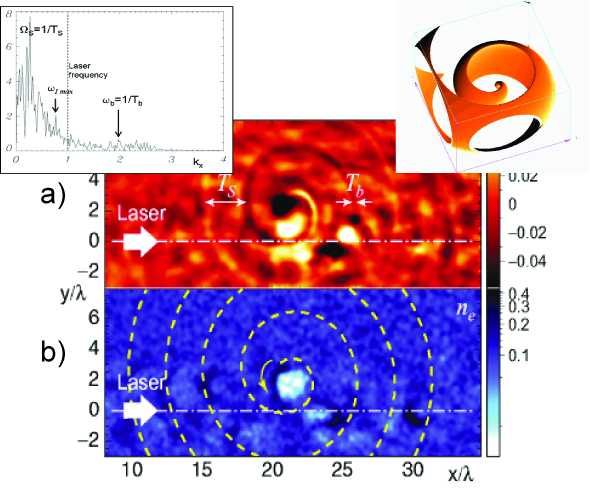} }
\caption{ a) Cross sections of the magnetic field component $eBz=m_ec$ in
the plane $x, y$. b) The electron density distribution. Right inset: The EM
field of the rotating electric charge rotating. Left inset: A frequency
spectrum of the emitted EM wave.}
\label{fig:16}
\end{figure}

The 3D solitons emit high-frequency EM radiation, whose frequency is much
higher than the Langmuir frequency \cite{Afterglow}. This radiation is
emanated from the electron density hump rotating in the wall of the soliton
cavity, similar to coherent synchrotron-like emission. This radiation has
the characteristics of a well pronounced outgoing spiral EM wave, Fig. \ref%
{fig:16} a). The emission of the spiral wave correlates to the rotation of
the electron density hump in the cavity wall, and it leads to the spiral
modulations of the electron density (see Fig. \ref{fig:16} b)). The density
hump gyrates in a circle, and the period of revolution is exactly equal to
the soliton period. The polarization of the spiral wave corresponds to the
well known synchrotron radiation \cite{VLG} and the density hump emission is
coherent. 

The results of 3D PIC simulations, presented in Ref. \cite{Afterglow},
distinctly demonstrated relativistic rotating dipoles excited by the
circularly polarized laser pulse in an underdense plasma. The dipoles are
associated with the relativistic electromagnetic solitons.

\section{Shock Waves}

\label{Sec:SW}

Phenomena taking place at shock-wave fronts play a key role in various
astrophysical conditions. The characteristic dimensionless parameters that
determine the shock wave propagagation are the magnetic Mach number, $%
M_{A}=v_{SW}/v_{A}$, equal to the ratio of the shock wave front velocity, $%
v_{SW}$, to the Alfven velocity, $v_{A}$, the ratio of the gas pressure to
the magnetic pressure, $8\pi nT/B^{2}$, and $\theta $, the angle between the
normal to the front and the magnetic field.

\begin{figure}[tbp]
\resizebox{0.4\textwidth}{!}{\includegraphics{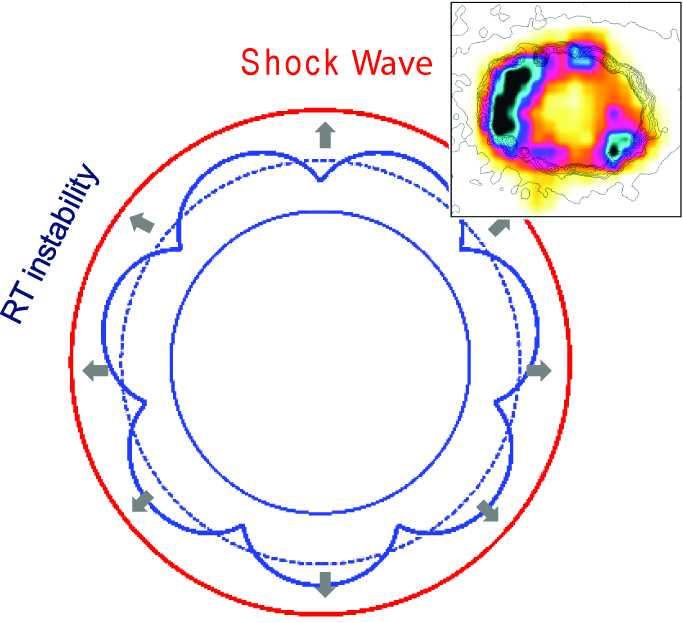} }
\caption{ Schematic view of the shock waves and contact dicontinuity in a
supernova remnant. In the inset: X-ray and optical image of supernova
remnant SNR 1987A \protect\cite{DBurrows,ImNa}.}
\label{fig:17}
\end{figure}

\subsection{Shock Waves in Supernova Remnants}

The origin of cosmic rays (CR) is one of the most interesting problems in
astroparticle physics \cite{ACR,CRLEC}. The observation of ultrahigh-energy
cosmic rays indicates that cosmic rays exist beyond 10$^{20}$ eV and
certainly beyond 10$^{19}$ eV energies greater than the GZK cutoff \cite{GZK}
for the extragalactic sources\ due to the pionization loss of protons that
decay by collision with cosmic microwave background photons. The galactic CR
spectra in the energy range above a few GeV and below $\approx 10^{7}$GeV
are power-laws with the total cosmic ray spectrum being 
\begin{equation}
I_{CR}=1.8\times \mathcal{E}^{-\kappa }\frac{\mathrm{particles}}{\mathrm{cm}%
^{2}\mathrm{s\ st\ GeV}}  \label{eq:CRspectr}
\end{equation}%
in the energy range from a few GeV to 100 TeV with $\kappa \approx 2.7$.
Around 10$^{15}$ eV (the \textquotedblleft knee\textquotedblright ), the
slope steepens from $\kappa \approx 2.7$ to $\kappa \approx 3$. The energies
10$^{18}$ eV correspond to the ultra high energy cosmic rays (UHECR), which
sources are associated with active galactic nuclei (AGNs) \cite{AUGER}.

For the most advanced theoretical models of galactic cosmic ray acceleration
with the energy below 10$^{17}$ eV the shock waves formed in the supernova
explosions are most important. This process is related to the nature of
collisionless shock waves \cite{TKrll}.

During explosions of type II supernovae an energy $\mathcal{E}_{tot}$ of the
order of $10^{51}$erg is released. The frequency of supernova explosions is
about 1/30 per year. Estimates \cite{ACR} show that approximately 2\% of the
energy of a supernova should be transferred into the cosmic ray energy.

In the initial stage of the evolution of a supernova envelope a system of
shocks is formed (\ref{fig:17}). The matter ejected from a star is
decelerated and compressed in the inner shock wave. Through the
circumstellar gas a second shock wave propagates. The matter ejected from a
star is separated from the circumstellar gas by a contact discontinuity,
which is unstable with respect to a Rayleigh-Taylor instability. The RT
instability leads to the relatively long scale modulations of the gas
density inside the supernova shells.

When the mass of the swept interstellar gas becomes larger than the mass
ejected from the star, the propagation of outer shock in Fig. (\ref{fig:17})
is described by the Sedov-Taylor self-similar solution. The radius of the
shock, $R_{SW}$, as a function of time is related to the energy, $\mathcal{E}%
_{SN}$, released in the explosion and to the gas density $\rho _{0}$ by the
relation 
\begin{equation}
R_{SW}(t)=1.51\left( \frac{\mathcal{E}_{SN}}{\rho _{0}}\right) ^{1/5}t^{2/5}=%
\frac{5}{2}v_{SW}t.
\end{equation}%
The shock wave velocity, $v_{SW}(t)\approx t^{-3/5}$, decreases with time.
At a later time when the radiation losses become important, the law of the
supernova envelope expansion changes. The asymptotic time dependence of the
SN envelope radius is given by $R_{SW}(t)\approx t^{2/7}$ (see Ref. \cite%
{ACR} and references therein).

\subsection{Collisionless Shock Waves}

If the shock wave has a relatively small amplitude, $M_{A}<M_{1}\approx 1.5$
(the precise value depends on $\beta $ and $\theta $), then the front
profile is laminar in structure and it is determined by a joint action of
the dispersion and dissipation on the nonlinear waves propagation. These
effects are described in the framework of the Korteweg-de Veries-Burgers
equation: 
\begin{equation}
\partial _{t}u+u\partial _{x}u-\nu \partial _{xx}u+\beta \partial _{xxx}u=0.
\label{eq:10kdvB}
\end{equation}

The stationary wave propagating with constant velocity is described by a
solution, which shows the change of the amplitude of the wave from zero far
ahead of the shock wave front, to $u_{1}=2v_{sw}$ far behind the shock wave
front.

\begin{figure}[tbp]
\resizebox{0.45\textwidth}{!}{\includegraphics{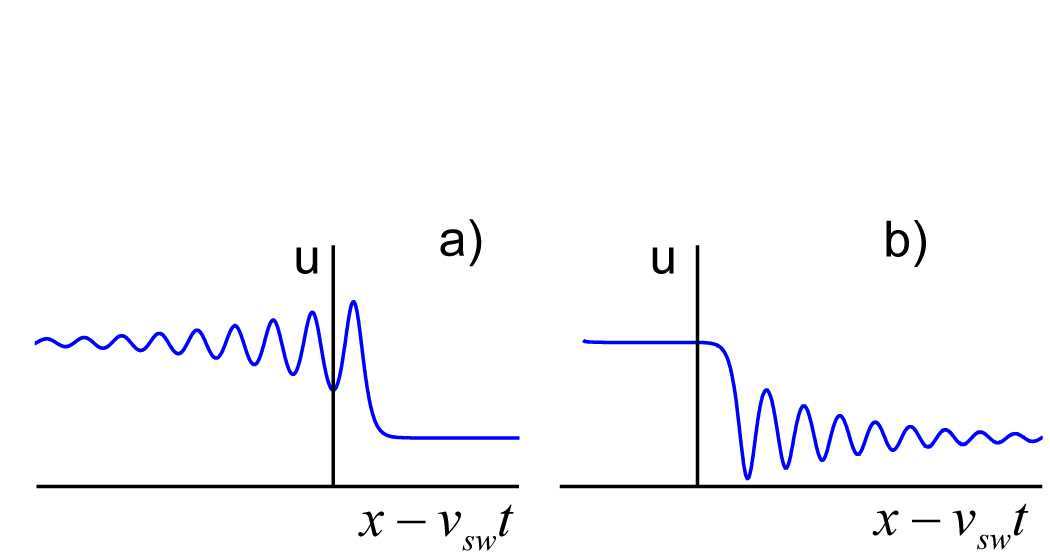} }
\caption{Structure of collisionless shock wave front for $\protect\beta >0$
(a) and for $\protect\beta <0$ (b).}
\label{fig:18}
\end{figure}

The decay of the oscillation amplitude, with the coefficient equal $\nu
/\beta $, results in the decrease of the amplitude of solitons as it is
shown in Fig. \ref{fig:18}. 
If dissipation effects are more important than the effects of dispersion, $%
\nu /\beta \gg 1$, there are no oscillations at the shock wave front. More
precisely, the decay should be large enough, $\nu \gg v_{cr},$ with 
\begin{equation}
\nu _{cr}=\sqrt{4\beta u_{1}}.
\end{equation}
In this case the wave has a monotonous structure.

In the case of $\nu /\beta \ll 1,$ the dispersion effects are dominant and
there are many well seen solitons near the front. For $\beta >0$ the
oscillations are localized behind the front (Fig. \ref{fig:12.4}a), while
for $\beta <0$ they are ahead of the front (Fig.\ref{fig:12.4}b).

For example, in the case of the magnetoacoustic shock waves in a plasma,
propagating almost perpendicularly to the direction of the magnetic field,
the dispersion coefficient, $\beta \approx v_{a}c^{2}/2\omega _{pe}^{2}$, is
positive. This means that the oscillations are localized behind the front of
the magnetoacoustic shock wave propagating perpendicularly to the magnetic
field. When the direction of the magnetoacoustic wave propagation is almost
parallel to the direction of the magnetic field, the coefficient $\beta
\approx -v_{a}c^{2}/2\omega _{pi}^{2}$ is negative with $\omega _{pi}=\sqrt{%
4\pi ne^{2}/m_{i}}$. The oscillations at the front of the magnetoacoustic
shock wave, propagating quasi-parallely with respect to the magnetic field,
are localized ahead of the front.

Dissipation, which determines the distance of the oscillation decay, can be
due to anomalous resistance and viscosity arising from an excitation of the
plasma instability, i.e. the Weibel instability of counterpenetrating
plasmas. If the amplitude of the shock wave is large, $M_{A}>3$, a high
level of turbulent fluctuations of electric and magnetic fields are excited
ahead and behind the wave front.

In the laser-plasma physics context, the observation of collisionless shocks
was reported by several authors \cite{SHW}, aiming to reproduce
astrophysical phenomena in small scale laboratories. However, in general
when the shocks were observed with optical probing techniques, the front
structure could hardly be resolved. In Ref. \cite{LRom} the propagation in a
rarefied plasma ($n_{e}<10^{15}cm^{-3}$) of collisionless shock waves being
excited following the interaction of a long ($L=470ps$) and intense ($%
I=10^{15}Wcm^{-2}$) laser pulse with solid targets, has been investigated
via proton probing techniques \cite{PrIm}. The shocks' structures and
related electric field distributions were reconstructed with high spatial
and temporal resolution. The experimental results are described within the
framework of the nonlinear wave description based on the Korteweg--de
Vries--Burgers equation (\ref{eq:10kdvB}).

\subsection{Diffusive Acceleration of Charged Particles at the Shock Wave
Front}

The charged particle interaction with fluctuations of the electric and
magnetic field in a turbulent plasma may result in particle scattering and
diffusion. When the shock wave propagates in a turbulent medium, an average
velocity of electromagnetic fluctuations is different in the regions ahead
and behind the shock front. Efficiently the particle appears to move between
semitransparent (due to diffusion) walls with a decreasing distance between
them. A model transport equation describing the particle convection,
diffusion and acceleration has a form \cite{ACR} 
\begin{equation*}
\partial _{t}\mathrm{\,}f+\mathrm{div}(\mathbf{u}\mathrm{\,}f-D\mathrm{\,}%
\nabla f)=
\end{equation*}
\begin{equation}
\frac{1}{p^{2}}\partial _{p}\left[ p^{2}\left( \frac{p}{3}\mathrm{div\ }%
\mathbf{u}-K(p)\right) \mathrm{\,}f\right] ,  \label{eq:CRtrEq}
\end{equation}
where $f(p,x,t)$ is the fast particle distribution function, $p$, $x$ and $t$
the particle momentum, coordinate and time, $v_{SW}$ being the speed of the
shock wave propagation, and $D$ is the diffusion coefficient. A term in the
right hand side describes regular acceleration or deceleration of the
charged particles: 
\begin{equation}
\frac{dp}{dt}=-K(p)-\frac{1}{3}p\ \mathrm{div\,}\mathbf{u}.
\end{equation}
The function $K(p)$ corresponds to the Compton and synchrotron losses
important for the cosmic ray electron component: 
\begin{equation}
K(p)=-\beta _{B}cp^{2}
\end{equation}
with 
\begin{equation}
\beta _{B}=8\times 10^{-25}\left( \frac{B^{2}}{8\pi }+w_{ph}\right) \frac{1}{%
eV\ s}
\end{equation}
An average change of the particle momentum proportional to $p\mathrm{\,div\,}%
\mathbf{u}$ occurs due to the particle bouncing between converging, $\mathrm{%
div\,}\mathbf{u}<0$ , or diverging, $\mathrm{div\,}\mathbf{u}<0$, scattering
centres.

\begin{figure}[tbp]
\resizebox{0.5\textwidth}{!}{\includegraphics{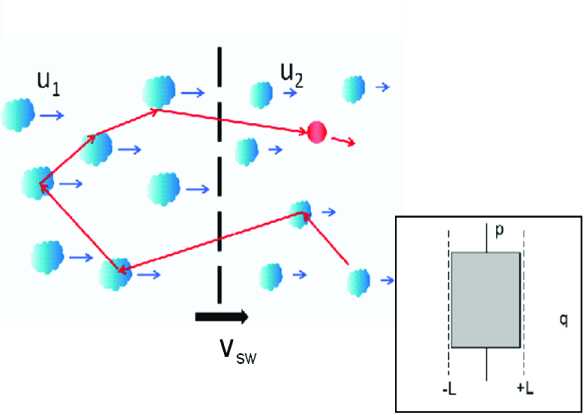} }
\caption{ Particle diffusion at the front of a shock wave propagating in a
turbulent plasma. Inset: the phase plane of the particle bouncing between
two plates. }
\label{fig:19}
\end{figure}

The average particle bouncing between two reflecting plates with distance $L 
$ \ as a function of time provides a simple example of a dynamic system with
conservation of the longitudinal adiabatic invariant, $J_{||}=pL$ \cite%
{LibLich,SVB-SNS}. The phase plane shown in the inset to Fig. \ref{fig:19},
illustrates the Fermi acceleration mechanism of the first type (type A
according to Ref. \cite{Fermi}). By virtue of the longitudinal adiabatic
invariant conservation, for decreasing distance between the plates, $dL/dt<0$, 
the particle momentum grows, i.e. the particle acquires energy.

The velocity distribution in the vicinity of the front of an infinitely thin
shock wave, propagating from left to right, has the form: $u(X)=u_{1}$ in
the region $X>0$, and $u(X)=u_{2}$ for $X<0$. Here $X=x-v_{SW}t$. The
velocities ahead the shock front and behind it are related to each other as 
\begin{equation}
u_{2}=u_{1}\frac{\kappa +1}{\kappa -1}. 
\end{equation}
Here, $\kappa $ is the polytropic
index. For an infinitely thin shock wave front the divergence of the
velocity is equal to 
\begin{equation}
\mathrm{div\ }\mathbf{u=(}u_{1}-u_{2})\delta (X).
\end{equation}

Substituting this expression into Eq. (\ref{eq:CRtrEq}), we obtain that the
charged particle acceleration at the fronts of collisionless shock waves
propagating in a turbulent plasma is described by the equation (see Ref. 
\cite{ACR} and references therein) 
\begin{equation*}
\partial _{X}(u(X)f-D\partial _{X}f)+\frac{1}{p^{2}}\partial
_{p}(p^{2}K(p)f)=
\end{equation*}%
\begin{equation}
-2\frac{u_{2}}{3(\kappa +1)}\delta (X)\frac{1}{p^{2}}\partial _{p}(p^{3}f).
\end{equation}%
In the limit, when the energy losses are negligibly small, this equation has
a solution, which gives a power law dependence of the distribution function, 
$f\propto p^{-k}$ with the index value $k=3u_{2}/(u_{2}-u_{1})$. For $\kappa
=5/3$ the index equals $k=4$, i.e. $f\propto p^{-4}$, or the energy spectrum 
$d\mathcal{N}_{CR}(\mathcal{E})/d\mathcal{E}\propto \mathcal{E}^{-3}$ is
close to the power law index observed in the galactic cosmic ray energy
spectrum (see Eq. (\ref{eq:CRspectr})).

For the cosmic ray electron component in the high energy limit, at the
energy when we cannot neglect the Compton and synchrotron losses, there is a
cut off in the spectrum \cite{SVBVAD}. For typical parameters in supernova
remnants, $D=10^{25}$cm$^{2}$s$^{-1}$, $B=10^{-4}$G, $u_{1}=10^{8}$ cm s$%
^{-1}$, the radiation losses limit the energy of ultrarelativistic electrons
by values of the order of 10 TeV.

Under the conditions of typical timescale of the laser plasmas the
synchrotron losses of ultrarelativistic electrons interacting with the
self-generated magnetic field is of the order of 
\begin{equation}
\tau _{B}=5\left( \frac{10^{3}}{\gamma _{e}}\right) \left( \frac{10^{9}G}{B}%
\right) ^{2}\ fs.
\end{equation}

\section{Conclusions}

Finally, we note that the development of superintense lasers with parameters
in the ELI range will provide the necessary conditions for experimental
physics where it will become possible to study ultrarelativistic energy of
accelerated charged particles, super high intensity EMW and the relativistic
plasma dynamics. A fundamental property of the plasma to create nonlinear
coherent structures, such as relativistic solitons and vortices,
collisionless shock waves and high energy particle beams, and to provide the
conditions for relativistic regimes of the magnetic field line reconnection,
makes the area of relativistic laser plasmas attractive for modeling of
processes of key importance for relativistic astrophysics.

\section*{Acknowledgement}

We appreciate discussions and comments from V. S. Beskin, M. Borghesi, P.
Chen, R. Diehl, A. Ya. Faenov, M. Kando, Y. Kato, T. Kawachi, J. K. Koga, K.
Kondo, G. Korn, G. Mourou, N. B. Narozhny, T. A. Pikuz, A. S. Pirozhkov, N.
N. Rosanov, V. I. Telnov, A. G. Zhidkov. This work was partially supported
by the Ministry of Education, Science, Sports and Culture of Japan,
Grant-in-Aid for Creative Scientific Research (A), 202244065, 2008.

The authors acknowledge the support by the European Commission under
contract ELI pp 212105 in the framework of the program FP7
Infrastructures-2007-1.

\end{document}